\documentstyle[psfig,amstex]{mn}

\def\etal{et~al.}
\def\spose#1{\hbox to 0pt{#1\hss}}
\def\lta{\mathrel{\spose{\lower 3pt\hbox{$\mathchar"218$}}
     \raise 2.0pt\hbox{$\mathchar"13C$}}}
\def\gta{\mathrel{\spose{\lower 3pt\hbox{$\mathchar"218$}}
     \raise 2.0pt\hbox{$\mathchar"13E$}}}

\def\clean{{\sc clean}}

\def\uv{{\it uv}}

\def\aips{{\sc aips}}

\def\imagr{{\sc imagr}}

\def\Ho50{$H_0 = 50$km\,s$^{-1}$\,Mpc$^{-1}$}

\title[Radio observations of 6C radio galaxies at $z \sim 1$]{Studies of a
sample of 6C radio galaxies at redshift one, I -- Deep multi--frequency
radio observations}  

\author[P.~N.~Best \etal]{P.~N.~Best,$^1$ S.~A.~Eales,$^2$
M.~S.~Longair,$^3$ S.~Rawlings$^4$ and H.~J.~A.~R\"ottgering$^1$\\
$^1$ Sterrewacht Leiden, Postbus 9513, 2300 RA Leiden, The Netherlands\\ 
$^2$ Department of Physics and Astrophysics, University of Wales Cardiff,
P.O.Box 913, Cardiff, CF2 3YB, United Kingdom\\
$^3$ Cavendish Astrophysics, Madingley Road, Cambridge, CB3 0HE, United
Kingdom\\
$^4$ Department of Astrophysics, University of Oxford, Keble Road, Oxford,
OX1 3RH, United Kingdom
} 

\pagerange{\pageref{firstpage}--\pageref{lastpage}}
\pubyear{1996}
\begin{document}
\label{firstpage}
\renewcommand{\dblfloatpagefraction}{0.9}
\renewcommand{\floatpagefraction}{0.9}
\renewcommand{\textfraction}{0.1}
\maketitle

\begin{abstract}
\noindent Deep radio observations at 5 and 8\,GHz are presented of a
complete sample of 11 radio galaxies with redshifts $0.85 < z < 1.5$,
selected from the 6C sample of Eales. The radio data, taken using the Very
Large Array in A, B and C array configurations, provide a best angular
resolution of 0.25 arcseconds and reach an rms noise level of order
20$\mu$Jy. Radio spectral index, radio polarisation, and rotation measure
maps are also presented for each source, and the radio data are compared
to K--band infrared images of the fields of these sources.

Radio core candidates are detected in eight of the eleven sources. Nine of
the eleven sources display deviations from `standard double radio source'
morphologies, with multiple hotspots in one or both lobes, or a hotspot
withdrawn from the leading edge of the radio emission. At 8\,GHz, the
sources are typically polarised at the 5 to 15\% level. The mean rotation
measures of the individual lobes are, in all but one case, less than
50\,rad\,m$^{-2}$, but strong asymmetries between the two lobes and steep
gradients within some lobes indicate that the Faraday rotation does not
have a Milky Way origin; rather, the distant 6C radio sources lie in a
relatively dense clumpy environment.

The sources are compared with the more radio powerful 3CR radio galaxies
in the same redshift range, and with low redshift radio galaxies. The
ratio of core to extended radio flux is found to be almost independent of
the linear size of a radio source and only weakly inversely correlated
with the total radio source power. This latter result indicates that the
high radio luminosity of the most powerful radio sources must originate in
a powerful active nucleus, in contrast to the suggestion of some authors
that such sources are so luminous only because confinement by a dense
surrounding environment boosts the lobe fluxes.  Environmental effects
must play a secondary role.
\end{abstract}

\begin{keywords}
Galaxies: active --- Radio continuum: galaxies --- Infrared: galaxies ---
Galaxies: jets --- Polarisation
\end{keywords}

\section{Introduction}

The revised 3CR radio source catalogue, defined by Laing, Riley and
Longair \shortcite{lai83}, consists of the most powerful radio galaxies in
the northern sky, selected at 178\,MHz. The revised 3CR sample has long
been fully identified optically, and is now also 100\% spectroscopically
complete.  Until recently, no other low--frequency selected radio source
sample even approached spectroscopic completeness.

The distant ($z \sim 1$) 3CR radio sources have been the targets of
numerous studies at a wide range of wavelengths (see e.g. McCarthy 1993
for a review).\nocite{mcc93} Of particular importance have been questions
as to the nature of the host galaxies of these radio sources, the age of
their stellar populations, the environment in which the galaxies live, and
the origin of their extremely powerful radio emission. These studies have
led to a number of surprising results, not least of which were the very
tight relationship between the infrared K--magnitudes and the redshifts of
these galaxies (e.g. Lilly and Longair 1984)\nocite{lil84a}, and the
discovery that the optical and ultraviolet emission of these galaxies is
elongated and aligned along the direction of the radio axis
\cite{cha87,mcc87}. Discerning the dependence of these properties upon
the radio power of the radio source is critical for determining the true
nature of these objects.

We are involved in a long--running programme to understand the
astrophysics of a sample of 28 3CR radio galaxies at redshift $z \sim 1$,
using observations in the optical wavebands with the Hubble Space
Telescope (HST), at radio wavelengths with the Very Large Array (VLA), and
in the near--infrared using the United Kingdom InfraRed Telescope (UKIRT)
\cite{lon95,bes96a,bes97c,bes98d}. In order to understand how these 3CR
radio galaxies relate to the less powerful radio galaxies, we have begun a
project to study a matched sample of eleven galaxies selected from the
6C/B2 sample of radio galaxies \cite{eal85} over a similar redshift
range. At any given redshift, these radio sources are about a factor of
six lower in radio luminosity than those selected from the 3CR catalogue.
Spectroscopic redshifts are currently available for 98\% of the 6C/B2
sample \cite{raw98}, making this an ideal sample for comparison with the
3CR radio galaxies.

In this paper we present deep radio observations at 8 and 5\,GHz of these
galaxies.  In Section~\ref{obsred}, we discuss the selection of the
current sample, the radio and infrared observations, and the data
reduction. The results of the current work are presented in
Section~\ref{images}, in the form of maps and images of the radio emission
of the sources, and its polarisation properties, at the various
frequencies. The properties of the radio sources are also tabulated.
Infrared K--band images of the fields of all but one of the radio
galaxies, published by Eales et~al \shortcite{eal97}, are compared with
the radio maps. In Section~\ref{discuss} we compare the 6C sources to
other radio galaxy samples, and our results are summarised in
Section~\ref{concs}. Presentation of the Hubble Space Telescope images of
the 6C galaxies and a discussion of the differences between the
optical\,/\,ultraviolet properties of the two samples is deferred to a
later paper (Best \etal\ in preparation, hereafter Paper II).

\section{Observations and data reduction}
\label{obsred}

\subsection {The sample}
\label{sample}

The current sample of radio galaxies was drawn from the complete sample of
59 radio sources from the 6CER sample \cite{raw98}, a revised version of
the sample originally defined by Eales \shortcite{eal85}. These radio
sources have flux densities at 151\,MHz which fall in the range 2.0\,Jy $<
S_{\rm 151} < 3.93$\,Jy, and lie in the region of the sky $08^{h} 20^{m} <
{\rm RA} < 13^{h} 01^{m}$, $34^{\circ} < {\rm Dec} < 40^{\circ}$.  Our
sample was restricted to those sources identified as radio galaxies and
with redshifts in the range $0.85 < z < 1.5$.  1212+38, whose redshift has
recently been revised downwards and now falls within our redshift range,
is not included in our subsample because at the time the project was begun
its redshift fell outside our selection criteria. 1123+34 was excluded
from the subsample due to its small angular extent ($\approx 0.5$ arcsec),
which would make it barely resolvable in the 5\,GHz observations. This
leaves a sample of 11 radio galaxies.

\subsection {Very Large Array observations}
\label{vla}

Observations of all eleven radio galaxies were made at 8\,GHz and 5\,GHz
using the A--array configuration of the VLA on December 8th, 9th and 10th,
1996. For ten of the sources a 50\,MHz bandwidth was used at each
frequency, but for the largest source, 1011+36, a 25\,MHz bandwidth was
used for the 8\,GHz observations to avoid chromatic aberration
effects. The largest angular scales that can be imaged using the A--array
of the VLA at 5 and 8\,GHz are about 11 and 8 arcsec respectively. Sources
larger than these sizes were also imaged using the B--array configuration;
these observations were made on February 24th, 1997. Similarly, the two
largest sources were imaged using the C--array configuration on September
17th 1997. Details of the observations are given in Table~\ref{vlatab}.

\begin{table}
\caption{\label{vlatab} Details of the VLA observations}
\begin{center}
\begin{tabular}{lccccr}
Source & $z$  &Frequencies& Array & Observing  & Int.\hspace{0.8mm} \\
       &      &           &config.&  Date      & time\hspace{0.4mm} \\
       &      &  [MHz]    &       &            & [min]\hspace{0.0mm} \\ 
0825+34& 1.46 & 4535,4885 &  A    & 10/12/96   & 39\hspace{1.0mm} \\
       &      & 8085,8335 &  A    & 09/12/96   & 103\hspace{1.0mm} \\
0943+39& 1.04 & 4535,4885 &  A    & 10/12/96   & 39\hspace{1.0mm} \\
       &      & 8085,8335 &  A    & 08/12/96   & 105\hspace{1.0mm} \\   
       &      &           &  B    & 24/02/97   & 60\hspace{1.0mm} \\
1011+36& 1.04 & 4535,4885 &  A    & 10/12/96   & 38\hspace{1.0mm} \\
       &      &           &  B    & 24/02/97   & 20\hspace{1.0mm} \\
       &      &           &  C    & 17/09/97   & 12\hspace{1.0mm} \\
       &      & 8085,8335 &  A    & 08/12/96   & 170\hspace{1.0mm} \\
       &      &           &  B    & 24/02/97   & 61\hspace{1.0mm} \\
       &      &           &  C    & 17/09/97   & 33\hspace{1.0mm} \\
1017+37& 1.05 & 4535,4885 &  A    & 10/12/96   & 39\hspace{1.0mm} \\
       &      & 8085,8335 &  A    & 09/12/96   & 91\hspace{1.0mm} \\
1019+39& 0.92 & 4535,4885 &  A    & 10/12/96   & 39\hspace{1.0mm} \\
       &      & 8085,8335 &  A    & 10/12/96   & 100\hspace{1.0mm} \\
1100+35& 1.44 & 4535,4885 &  A    & 10/12/96   & 39\hspace{1.0mm} \\
       &      &           &  B    & 24/02/97   & 21\hspace{1.0mm} \\
       &      & 8085,8335 &  A    & 08/12/96   & 70\hspace{1.0mm} \\
       &      &           &  A    & 09/12/96   & 33\hspace{1.0mm} \\
       &      &           &  B    & 24/02/97   & 61\hspace{1.0mm} \\
1129+37& 1.06 & 4535,4885 &  A    & 10/12/96   & 39\hspace{1.0mm} \\
       &      &           &  B    & 24/02/97   & 21\hspace{1.0mm} \\
       &      & 8085,8335 &  A    & 09/12/96   & 105\hspace{1.0mm} \\
       &      &           &  B    & 24/02/97   & 61\hspace{1.0mm} \\
1204+35& 1.37 & 4535,4885 &  A    & 10/12/96   & 38\hspace{1.0mm} \\
       &      & 8085,8335 &  A    & 09/12/96   & 104\hspace{1.0mm} \\
       &      &           &  B    & 24/02/97   & 60\hspace{1.0mm} \\
1217+36& 1.09 & 4535,4885 &  A    & 10/12/96   & 39\hspace{1.0mm} \\
       &      & 8085,8335 &  A    & 08/12/96   & 105\hspace{1.0mm} \\
1256+36& 1.07 & 4535,4885 &  A    & 10/12/96   & 39\hspace{1.0mm} \\
       &      &           &  B    & 24/02/97   & 21\hspace{1.0mm} \\
       &      & 8085,8335 &  A    & 08/12/96   & 105\hspace{1.0mm} \\
       &      &           &  B    & 24/02/97   & 62\hspace{1.0mm} \\
1257+36& 1.00 & 4535,4885 &  A    & 10/12/96   & 39\hspace{1.0mm} \\
       &      &           &  B    & 24/02/97   & 22\hspace{1.0mm} \\
       &      &           &  C    & 17/09/97   & 11\hspace{1.0mm} \\
       &      & 8085,8335 &  A    & 09/12/96   & 105\hspace{1.0mm} \\
       &      &           &  B    & 24/02/97   & 62\hspace{1.0mm} \\
       &      &           &  C    & 17/09/97   & 33\hspace{1.0mm} \\
\end{tabular}
\end{center}
\end{table}

The observations were carried out using standard VLA procedures. Short
observations of the primary flux calibrator 3C286 were used to calibrate
the flux density scale, and observations of this source separated in time
by 6 hours determined the absolute polarisation position angle. The
uncertainty in the calibration of the position angles, estimated from the
difference between the solutions for the scans of 3C286, was about $\pm$2
degrees at each frequency. For observations at 4710 and 8210\,MHz this
corresponds to an uncertainty in the absolute value of the rotation
measure of about 20\,rad\,m$^{-2}$. Secondary calibrators within a few
degrees of each source were observed approximately every 25 minutes to
provide accurate phase calibration; observations of these calibrators were
spaced over a wide range of parallactic angles enabling the on-axis
antenna polarisation response terms to be determined.

The data were reduced using the \aips\ software provided by the National
Radio Astronomy Observatory, with the two IFs at each frequency being
reduced separately. The data from each different array configuration were
individually \clean ed using the \aips\ task \imagr, and then one or two
cycles of phase self--calibration were used to improve further the map
quality. The \uv\ data from the lower resolution array configurations were
self--calibrating with those of the highest resolution array, a combined
dataset was then produced, and a further cycle of phase self--calibration
was carried out.

\subsection{The radio maps}
\label{radmaps}

For each source, images were made at full angular resolution in the Stokes
parameters I, Q and U, at both 8 and 5\,GHz by \clean ing the final
datasets using the \aips\ task \imagr. The full--width--half--maxima
(FWHM) of the Gaussian restoring beams used for each map are provided in
the figure captions. The Stokes I images of the two IFs at each frequency
were combined to produce single total intensity maps at 8210 and at
4710\,MHz.

Spectral index, rotation measure, depolarisation measure and magnetic
field position angle maps of the sources were made following the method
outlined by Best \etal\ \shortcite{bes98a}. In brief, images of the 8\,GHz
data were made at the resolution of the 5\,GHz data by applying an upper
cut--off in the \uv\ data matching the longest baseline sampled at 5\,GHz,
together with \uv\ tapering to maintain the smooth coverage of the \uv\
plane. Using these matched resolution datasets, maps of the spectral
index, $\alpha$, (where $I_{\nu} \propto \nu^{-\alpha}$) were made in
regions of the images with surface brightnesses in excess of 5 times the
rms noise level at both frequencies.  Rotation measures were derived from
the polarisation position angles at the four frequencies 4535, 4885, 8085
and 8335\,MHz, in regions where the polarised intensity was detected in
excess of four times the noise level at all frequencies. The
depolarisation measure, $DM_{8.2}^{4.7}$, defined as the ratio of the
fractional polarisation at 4710\,MHz to that at 8210\,MHz, was determined
on a pixel by pixel basis, after correcting for Ricean bias in the total
polarised intensity images, for pixels in which the total polarised
intensity at both frequencies exceeding five times the noise level.

\begin{table*}
\caption{\label{radprops} Properties of the radio sources. Total fluxes
are measured from the lowest resolution data at each frequency, to ensure
that all of the large--scale structure is sampled. They are quoted to the
nearest mJy, as determined from these maps; these values are subject to
errors of up to a few per cent due to the limited accuracy of the absolute
calibration of the VLA.  The fractional polarisation at each frequency was
derived by dividing the flux density of the total polarised intensity map,
after correction for Ricean bias, by that of the total intensity map, and
is therefore a scalar rather than vector average of the polarisation. As
discussed in the text, the Ricean bias correction may also lead to null
polarisation measurements from low surface brightness regions which are
polarised, particularly for the largest sources; therefore the fractional
polarisations quoted should strictly be treated as lower limits, and no
estimate of the error in this measurement is made. The largest angular
size of a source is as measured between the centres of the compact
emission region in each lobe which is most distant from the AGN; this
measurement has an associated error $\lta 0.1''$. The separation quotient,
$Q$, is defined as the ratio of the angular separations of the hotspots in
the longer and shorter arms from the nucleus (tabulated in
Table~\ref{regprop}). The core fraction is calculated as the ratio of the
flux density of the compact central component to that of the extended
radio emission at 8\,GHz. Where a core candidate is not detected, an upper
limit for the core fraction is derived assuming the core flux density to
be below five times the rms noise level.  The beam sizes of the radio maps
are provided in the individual figure captions.}
\begin{center}
\begin{tabular}{lllrrcrrcrcc}
Source&\multicolumn{2}{c}{RA~~~~(J2000)~~~~Dec}&Total &
Frac.\hspace*{1mm} &RMS&Total &Frac.\hspace*{1mm} & 
RMS&Largest& $Q$ & Core \\
&&&Flux\hspace*{0.2mm} &Polaris.&Noise&Flux\hspace*{0.2mm} &
Polaris.&Noise&Angular&&Fraction \\
&&&\multicolumn{3}{c}{............8210\,MHz..........}&
\multicolumn{3}{c}{............4710\,MHz..........}&Size\hspace*{2.2mm} &&8210\,MHz\\
&&&[mJy]\hspace*{0.1mm} &[\%]\hspace*{3mm} &[$\mu$Jy]&[mJy]\hspace*{0.1mm} 
&[\%]\hspace*{3mm} &[$\mu$Jy]&[$''$]\hspace*{2.8mm}& &      \\
0825+34&08 28 26.85&34 42 49.1    & 45\hspace*{1.6mm} & 6.9\hspace*{3mm} 
& 18 & 87\hspace*{1.6mm} & 6.3\hspace*{3mm} & 22 &  7.0\hspace*{2.5mm} 
& $1.53 \pm 0.07$ & $0.0053 \pm 0.0004$ \\
0943+39&09 46 18.71&39 44 18.5    & 45\hspace*{1.6mm} & 7.7\hspace*{3mm} 
& 15 & 78\hspace*{1.6mm} & 2.7\hspace*{3mm} & 26 & 10.6\hspace*{2.5mm} 
& $3.44 \pm 0.05$ & $0.0049 \pm 0.0003$ \\
1011+36&10 14 12.90&36 17 18.0    & 52\hspace*{1.6mm} & 2.7\hspace*{3mm} 
& 19 & 87\hspace*{1.6mm} & 2.9\hspace*{3mm} & 26 & 51.0\hspace*{2.5mm} 
& $1.26 \pm 0.01$ & $0.0943 \pm 0.0005$ \\
1017+37&10 20 40.03&36 57 02.3    & 57\hspace*{1.6mm} & 8.0\hspace*{3mm} 
& 19 & 109\hspace*{1.6mm} & 6.7\hspace*{3mm} & 23 & 7.4\hspace*{2.5mm} 
& $3.48 \pm 0.21$ & $0.0021 \pm 0.0003$ \\
1019+39&10 22 55.25&39 08 49.9$^a$& 73\hspace*{1.6mm} &10.6\hspace*{3mm}  
& 20 & 139\hspace*{1.6mm} & 7.2\hspace*{3mm} & 24 &7.9\hspace*{2.5mm}  
& $1.3 \pm 0.4$ &$<$0.0014 \\
1100+35&11 03 26.26&34 49 47.2    & 56\hspace*{1.6mm} & 8.0\hspace*{3mm}
& 14 & 99\hspace*{1.6mm} & 7.5\hspace*{3mm} & 28 & 13.1\hspace*{2.5mm} 
& $1.40 \pm 0.03$ & $0.124 \pm 0.001$ \\
1129+37&11 32 35.35&36 54 17.8$^a$& 75\hspace*{1.6mm} &16.8\hspace*{3mm} 
& 14 & 134\hspace*{1.6mm} &15.3\hspace*{3mm} & 22 &16.1\hspace*{2.5mm} 
& $1.7 \pm 0.3$ &$<$0.0009 \\ 
1204+35&12 07 31.86&35 03 06.2    & 62\hspace*{1.6mm} &11.8\hspace*{3mm} 
& 14 & 117\hspace*{1.6mm} & 9.1\hspace*{3mm} & 27 &17.4\hspace*{2.5mm} 
& $1.55 \pm 0.03$ & $0.021 \pm 0.0002$ \\
1217+36&12 20 09.83&36 29 07.1    &111\hspace*{1.6mm} & 5.5\hspace*{3mm} 
& 23 & 163\hspace*{1.6mm} & 4.6\hspace*{3mm} & 45 & 4.3\hspace*{2.5mm} 
&  --- & $\lta 0.67^b$ \\   
1256+36&12 59 06.07&36 31 58.2$^a$& 97\hspace*{1.6mm} &10.6\hspace*{3mm} 
& 15 & 172\hspace*{1.6mm} &10.3\hspace*{3mm} & 25 &17.5\hspace*{2.5mm} 
& $1.25 \pm 0.20$ &$<$0.0008 \\
1257+36&12 59 30.02&36 17 03.0$^c$& 40\hspace*{1.6mm} & 7.8\hspace*{3mm} 
& 12 & 73\hspace*{1.6mm} & 7.1\hspace*{3mm} & 26 & 38.9\hspace*{2.5mm} 
& $ 1.14 \pm 0.01$ & $0.0032 \pm 0.0003^c$ \\
\end{tabular}
\end{center}
\raggedright $^a$ Positions taken from optical IDs.\\ \raggedright $^b$
Core fraction probably overestimated: see text.\\ \raggedright $^c$
Assuming that the core is the NW candidate. If it is the SE candidate then
the position is RA: 12 59 30.13, Dec: +36 17 02.0, and the core fraction
is $0.0055 \pm 0.0003$.
\end{table*}

\begin{table*}
\caption{\label{regprop} Properties of the various components of the radio
sources.  Total fluxes and fractional polarisations are as defined in
Table~\ref{radprops}. Unless explicitly stated, errors in the total flux
measurements are of order a few per cent due to calibration
uncertainties. Fractional polarisations, as discussed for
Table~\ref{radprops}, are strictly lower limits. The spectral indices are
the mean values for that region of the source, calculated between 4710 and
8210\,MHz; assuming 3\% uncertainties in the absolute calibration at each
frequency gives a systematic uncertainty in the absolute spectral indices
of 0.07. Uncertainties due to measurement errors of individual features
are quoted when comparable to or larger than this systematic error.  The
depolarisation measures quoted are the mean values of the pixel by pixel
ratios of the scalar fractional polarisation at 4710 and 8210\,MHz in each
region; in this way, they are unaffected by regions in which polarisation
is too low to be measured. The errors represent the error on the mean in
these regions. The rotation measures and their errors are the mean values
from the frequencies 4535, 4885, 8085 and 8335\,MHz determined in a
similar way.  The angular size quoted for each lobe is the angular
separation between the radio core and the most distant compact emission
region in the lobe. Where a radio core is detected these are measured to
an accuracy $\lta 0.1''$; in the three cases where the optical ID is used,
the accuracy is only $\sim 1$ arcsec.}
\begin{center}

\begin{tabular}{lccrcrrcrc}
Source&Component&Total &Fractional\hspace*{1mm} &Total 
&Fractional\hspace*{1mm} &Spectral\hspace*{2mm} &Depolar.&Rotation\hspace*{0.2mm} &Angular\\
&&Flux&Polarization&Flux &Polarization
&Index\hspace*{4mm} &Measure&Measure\hspace*{0.3mm} &Size \\
&&8210\,MHz&8210\,MHz\hspace*{1.5mm} &4710\,MHz&4710\,MHz\hspace*{1.5mm} 
&$\alpha$\hspace*{6.5mm} &$DM_{8.2}^{4.7}$\hspace*{0.3mm} &$RM$\hspace*{3mm} &\\
&&[mJy] &[\%]\hspace*{5mm} &[mJy] &
[\%]\hspace*{5mm} &&&[rad\,m$^{-2}$]&[$''$] \\
0825+34& Core    & $0.24 \pm 0.018$ &  0.0\hspace*{6mm} & $0.16 \pm 0.022$ 
& 0.0\hspace*{6mm} &   $-0.73 \pm 0.28$  & ---  & ---\hspace*{4mm} & --- \\
       & SE lobe &        2.6       &  1.9\hspace*{6mm} &  4.6 
&  0.0\hspace*{6mm} &  1.04\hspace*{5mm} & 0.0  & ---\hspace*{4mm} & 4.2 \\   
       & NW lobe &       42.4       &  7.1\hspace*{6mm} & 82.1 
&  6.6\hspace*{6mm} &  1.19\hspace*{5mm} & $0.94 \pm 0.04$& $14 \pm 2$ \hspace*{1mm}   & 2.75\\ 
0943+39& Core    & $0.22 \pm 0.015$ &  0.0\hspace*{6mm} & $0.20 \pm 0.026$ 
&  0.0\hspace*{6mm} & $-0.18 \pm 0.26$   &---  & ---\hspace*{4mm} & --- \\ 
       & E lobe  &       21.9       &  9.2\hspace*{6mm} & 44.3 
&  1.4\hspace*{6mm} &  1.27\hspace*{5mm} & $0.22 \pm 0.01$& $26 \pm 2$ \hspace*{1mm}  & 2.4 \\  
       & W lobe  &       22.6       &  6.5\hspace*{6mm} & 33.8 
&  4.4\hspace*{6mm} &  0.73\hspace*{5mm} & $0.84 \pm 0.02$& $12 \pm 2$ \hspace*{1mm}  & 8.25\\
1011+36& Core    &       4.9        &  0.0\hspace*{6mm} &  4.0 
&  0.0\hspace*{6mm}&$-$0.33\hspace*{5mm} &---  & ---\hspace*{4mm} & --- \\ 
       & N lobe  &       32.9       &  3.1\hspace*{6mm} & 56.9 
&  3.1\hspace*{6mm} &  0.99\hspace*{5mm} & $1.02 \pm 0.02$& $41 \pm 2$ \hspace*{1mm}   & 22.8\\
       & S lobe  &       13.9       &  2.7\hspace*{6mm} & 26.2 
&  2.9\hspace*{6mm} &  1.15\hspace*{5mm} & $0.91 \pm 0.02$& $15 \pm 1$ \hspace*{1mm} & 28.8\\
1017+37& Core    & $0.12 \pm 0.019$ &  0.0\hspace*{6mm} & $0.10 \pm 0.023$
&  0.0\hspace*{6mm} & $-0.24 \pm 0.50$   &---  & ---\hspace*{4mm} & --- \\ 
       & NE lobe &       34.4       &  6.9\hspace*{6mm} & 66.7 
&  5.8\hspace*{6mm} &  1.19\hspace*{5mm} & $0.78 \pm 0.07$&$-13 \pm 5$ \hspace*{1mm} &1.65\\ 
       & SW lobe &       22.4       &  9.6\hspace*{6mm} & 40.1 
&  8.3\hspace*{6mm} &  1.04\hspace*{5mm} & $0.87 \pm 0.02$& $22 \pm 3$ \hspace*{1mm}  & 5.75\\
1019+39& NE lobe &       28.7       & 11.6\hspace*{6mm} & 54.2 
&  8.5\hspace*{6mm} &  1.14\hspace*{5mm} & $0.88 \pm 0.01$& $38 \pm 3$ \hspace*{1mm}  &4.5$^a$\\
       & SW lobe &       45.1       &  9.6\hspace*{6mm} & 84.9 
&  6.4\hspace*{6mm} &  1.14\hspace*{5mm} & $0.81 \pm 0.04$& $89 \pm 4$ \hspace*{1mm}  &3.5$^a$\\
1100+35& Core    &        6.9       &  0.0\hspace*{6mm} &  6.8 
&  0.0\hspace*{6mm} &$-$0.04\hspace*{5mm}&---  & ---\hspace*{4mm} & --- \\
       & `Jet'   &        7.0       &  6.8\hspace*{6mm} & 11.1 
&  4.3\hspace*{6mm} &  0.82\hspace*{5mm} & $0.86 \pm 0.02$& $23 \pm 2$ \hspace*{1mm}  & --- \\ 
       & E lobe  &       12.7       &  9.7\hspace*{6mm} & 21.7 
&  9.6\hspace*{6mm} &  0.96\hspace*{5mm} & $1.01 \pm 0.02$& $10 \pm 3$ \hspace*{1mm}  & 7.7 \\
       & W lobe  &       29.1       &  9.5\hspace*{6mm} & 57.0 
&  8.5\hspace*{6mm} &  1.21\hspace*{5mm} & $0.91 \pm 0.01$& $17 \pm 1$ \hspace*{1mm}  & 5.5 \\
1129+37& NW lobe &       30.0       &  9.6\hspace*{6mm} & 52.8 
&  8.3\hspace*{6mm} &  1.02\hspace*{5mm} & $0.91 \pm 0.02$& $-13 \pm 2$ \hspace*{1mm} &6$^a$\\
       & SE lobe &       45.5       & 21.3\hspace*{6mm} & 80.9 
& 19.9\hspace*{6mm} &  1.03\hspace*{5mm} & $0.93 \pm 0.01$& $3 \pm 2$  \hspace*{1mm}  &10$^a$\\
1204+35& Core    & $1.33 \pm 0.014$ &  0.0\hspace*{6mm} & $2.19 \pm 0.027$ 
&  0.0\hspace*{6mm} &  0.90\hspace*{5mm} & ---  & ---\hspace*{4mm} & --- \\
       & N lobe  &       27.5       & 17.8\hspace*{6mm} & 53.2 
& 14.0\hspace*{6mm} &  1.19\hspace*{5mm} & $0.84 \pm 0.02$& $-4 \pm 1$ \hspace*{1mm} & 6.85\\
       & S lobe  &       32.8       &  7.3\hspace*{6mm} & 62.2 
&  5.2\hspace*{6mm} &  1.15\hspace*{5mm} & $0.88 \pm 0.03$ & $-24 \pm 3$ \hspace*{1mm} &10.6\\ 
1217+36& `Core'  &       74.3       &  4.6\hspace*{6mm} & 92.8 
&  5.4\hspace*{6mm} &  0.40\hspace*{5mm} & $1.13 \pm 0.02$& $10 \pm 2$ \hspace*{1mm}  & --- \\
       & Diffuse &       36.4       &  7.4\hspace*{6mm} & 70.0 
&  3.5\hspace*{6mm} &  1.18\hspace*{5mm} & $0.72 \pm 0.04$& $-9 \pm 5$ \hspace*{1mm}  & --- \\
1256+36& NE lobe &       35.8       &  9.9\hspace*{6mm} & 63.6 
& 10.3\hspace*{6mm} &  1.03\hspace*{5mm} & $1.03 \pm 0.03$ & $-13 \pm 2$ \hspace*{1mm} &10$^a$\\
       & SW lobe &       60.5       & 11.2\hspace*{6mm} &108.2 
&  9.9\hspace*{6mm} &  1.04\hspace*{5mm} & $0.88 \pm 0.01$ & $-3 \pm 1$ \hspace*{1mm} & 8$^a$\\
1257+36&`Core' NW&  $0.13 \pm 0.012$ &  0.0\hspace*{6mm} & $0.11 \pm 0.026$ 
&  0.0\hspace*{6mm} & $-0.31 \pm 0.46$   &---  & ---\hspace*{4mm} & --- \\
       &`Core' SE&  $0.22 \pm 0.012$ &  0.0\hspace*{6mm} & $0.30 \pm 0.026$ 
&  0.0\hspace*{6mm} & $ 0.55 \pm 0.18$ & ---  & ---\hspace*{4mm} & --- \\
       & SE lobe &       15.9       &  2.3\hspace*{6mm} & 28.7 
&  1.9\hspace*{6mm} &  1.06\hspace*{5mm} & $0.99 \pm 0.03$& $35 \pm 2$ \hspace*{1mm} & 20.8\\
       & NW lobe &       22.9       & 12.0\hspace*{6mm} & 42.4 
& 10.9\hspace*{6mm} &  1.11\hspace*{5mm} & $0.96 \pm 0.01$ & $9 \pm 1$ \hspace*{1mm} & 18.2\\
\end{tabular}
\end{center}
\raggedright $^a$ Based upon positions of optical IDs.
\end{table*}

\subsection{The infrared data}
\bigskip

Infrared K--band images of all of these galaxies except 1019+39 have been
taken during three observing runs, using the IRCAM and IRCAM3 cameras on
the United Kingdom Infrared Telescope (UKIRT) and the REDEYE camera on the
Canada--France--Hawaii Telescope (CFHT). These data were presented and
described by Eales \etal\ \shortcite{eal97}. Here, we present a comparison
of this K--band data with the new radio data.

Uncertainties in the relative alignment of the infrared and radio
reference frames may result in astrometric errors between the radio and
infrared images of about an arcsecond: the errors are largest when no
fiducial stars are present within the small field of view of the infrared
cameras. For the eight galaxies for which a core candidate was detected
$\lta 1$ arcsec from the infrared galaxy, the relative alignment of the
two frames was improved by assuming the radio core to be co-incident with
the centroid of the infrared emission. Such alignment was possible to an
accuracy of about 0.2 arcsec. For the two galaxies in which no radio core
was identified, the infrared and radio images were simply overlaid
assuming the two reference frames to be accurately registered.

\section{Results}
\label{images}

In Figures 1 to 11, maps of the radio data are provided for each
source. Shown in each figure are the 8\,GHz radio map made at the highest
angular resolution, the 5\,GHz radio map, the polarisation position angle
of the electric field vectors at 8\,GHz, the magnetic field direction
determined wherever a rotation measure could be derived, greyscale plots
of the spectral index, the rotation measure and the depolarisation
measure, calculated as described in the previous section, and finally
(except for 1019+39) the infrared K--band image of the field of each radio
source. Important parameters of each source are provided in
Table~\ref{radprops}, and are determined for the various components of the
source in Table~\ref{regprop}.

Some words of caution should be added here concerning the interpretation
of the greyscale figures for spectral index, depolarisation and rotation
measures, and of the fractional polarisations given in Tables~2 and 3. The
inability of the \clean\ procedure to accurately represent smooth extended
low--surface brightness emission can lead to artefacts in the spectral
index maps, with such regions appearing speckled. This is particularly
noticeable for 1011+36, 1257+36, and the western lobe of 1100+35
(Figures~3e, 6e and 11e), but can be seen at fainter levels in other
sources. The globally averaged spectral indices of these regions,
presented in Table~3, consider all of the low surface brightness emission
regardless of how \clean\ has distributed it, and so are reliable. In
addition, polarised flux in regions of low surface brightness may be
removed by the Ricean bias correction, leading to these regions lacking
polarisation and rotation measure data; this means that the fractional
polarisations quoted in Tables~2 and 3 for sources containing such regions
could better be considered as lower limits, especially for the largest
sources. The depolarisation measures, however, are limited to regions
where polarised emission is measured at both frequencies, and so do give
accurate values. The global properties shown in the rotation measure
greyscales can be demonstrated to be fully reliable using fits of the
polarisation position angle against a $\lambda$--squared law; however,
individual large variations seen in areas of only 2--3 pixels,
particularly towards the extremities, should be treated with caution.

\subsection{Notes on individual radio sources}

A brief description of the structures of the individual radio sources
is provided below. 
\bigskip

${\it 0825+34:}$ This source shows a strong asymmetry in flux between its
two lobes, but the new detection of an inverted spectrum radio core
roughly co-incident with the optical identification of Eales \etal\
\shortcite{eal97} at redshift $z=1.46$ \cite{raw98}, demonstrates that
this source is an asymmetric double rather than a core--jet source
(c.f. Naundorf \etal\ 1992).\nocite{nau92} The south--eastern arm contains
double hotspots. The polarisation properties are strongly asymmetric, with
the north--western lobe having little depolarisation at 5\,GHz whilst the
south--eastern emission is completely depolarised.

\bigskip

${\it 0943+39:}$ A flat--spectrum radio core is detected for the first
time, co-incident with the proposed identification of a galaxy with
redshift $z=1.04$ \cite{eal97}. The eastern lobe shows very strong
depolarisation and, on the 8\,GHz image, the radio emission appears to
bend sharply at the location of the current hotspot. The western arm is
also depolarised, although much less so, and has a uniform, relatively
flat spectral index.

\bigskip

${\it 1011+36:}$ The largest source in the current sample, 1011+36 has a
bright, inverted--spectrum radio core, and is associated with a galaxy
with redshift $z=1.04$ \cite{eal97}, rather than the original
identification close to the northern radio lobe proposed by
Allington--Smith \etal\ (1982; see also Naundorf \etal\ 1992, Law--Green
\etal\ 1995).\nocite{all82b,nau92,law95} This original identification is
not even detected on our K--band image (Figure~3h). The image does,
however, show two faint companions within five arcsecs of the host galaxy,
and aligned close to the radio axis.

The south--western lobe of the source contains a second, compact
hotspot. In the 1.4\,GHz image of Law--Green \etal\, the lobe emission of
the northern arm extends back to the core, but its surface brightness at
5\,GHz is too low for this to be apparent on our map. The source shows low
polarisation, but practically no depolarisation between 5 and 8\,GHz. It
is possible that owing to the low surface brightness of many of the
features of this radio source, a significant fraction of the true
polarised intensity will have been lost during the Ricean bias removal.

\bigskip

${\it 1017+37:}$ First identified by Lilly \shortcite{lil89}, the radio
galaxy which hosts 1017+37 ($z=1.05$, Rawlings \etal\ 1998, also known as
4C37.27A)\nocite{raw98} lies close to the north--eastern lobe, and for the
first time a faint core candidate is detected co-incident with this. The
source shows a large asymmetry both in the angular sizes of its lobes, and
in their rotation measures; the rotation measures of the lobes differ by
nearly 150\,rad\,m$^{-2}$ in the rest--frame of the source.
 
${\it 1019+39:}$ Also known as 4C39.31, this radio source was identified
with a galaxy at redshift $z=0.921$ by Allington--Smith \etal\ (1985; see
also Thompson \etal\ 1994).\nocite{tho94,all85} No radio core is detected
in the current observations. The south--western lobe is the more compact,
and contains a second bright hotspot. The two radio lobes are reasonably
symmetric in their depolarisations and spectral indices, but differ by
about 200\,rad\,m$^{-2}$ in their rest--frame rotation measures.

\bigskip

${\it 1100+35:}$ The radio structure of this source strongly resembles
that of a quasar, with a luminous flat--spectrum core, a bright one--sided
jet leading to a compact lobe, and on the opposite side of the source a
much more diffuse lobe closer to the nucleus (see also Law--Green \etal\
1995)\nocite{law95}. Lilly \shortcite{lil89} omitted this from his sample
of radio galaxies on the basis of its bright infrared magnitude, and
Law--Green \etal\ classified it as a quasar. However, no broad lines are
seen in its optical spectrum (Rawlings \etal\ 1998; the signal--to--noise
ratio of the spectrum is low, however, and so the limits set are not
especially tight), and its infrared K--band emission is resolved
(Figure~6h). 1100+35 should therefore be classified as a galaxy. Although
strongly asymmetric in appearance and flux density, the two lobes of this
radio source show little difference in their polarisation properties.

\bigskip

\begin{figure*}
\centerline{
\psfig{figure=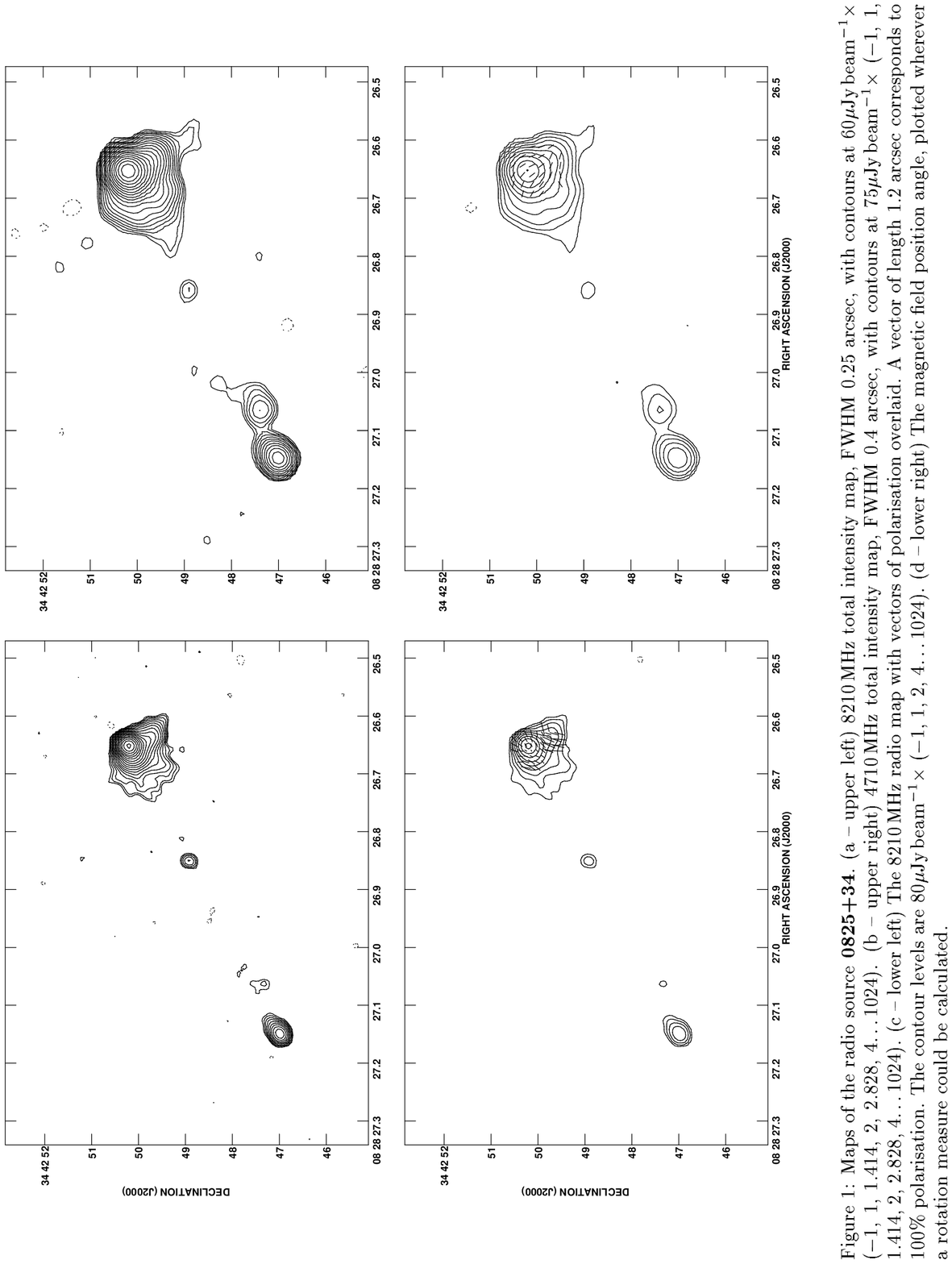,clip=,width=\textwidth}
}
\end{figure*}

\begin{figure*}
\centerline{
\psfig{figure=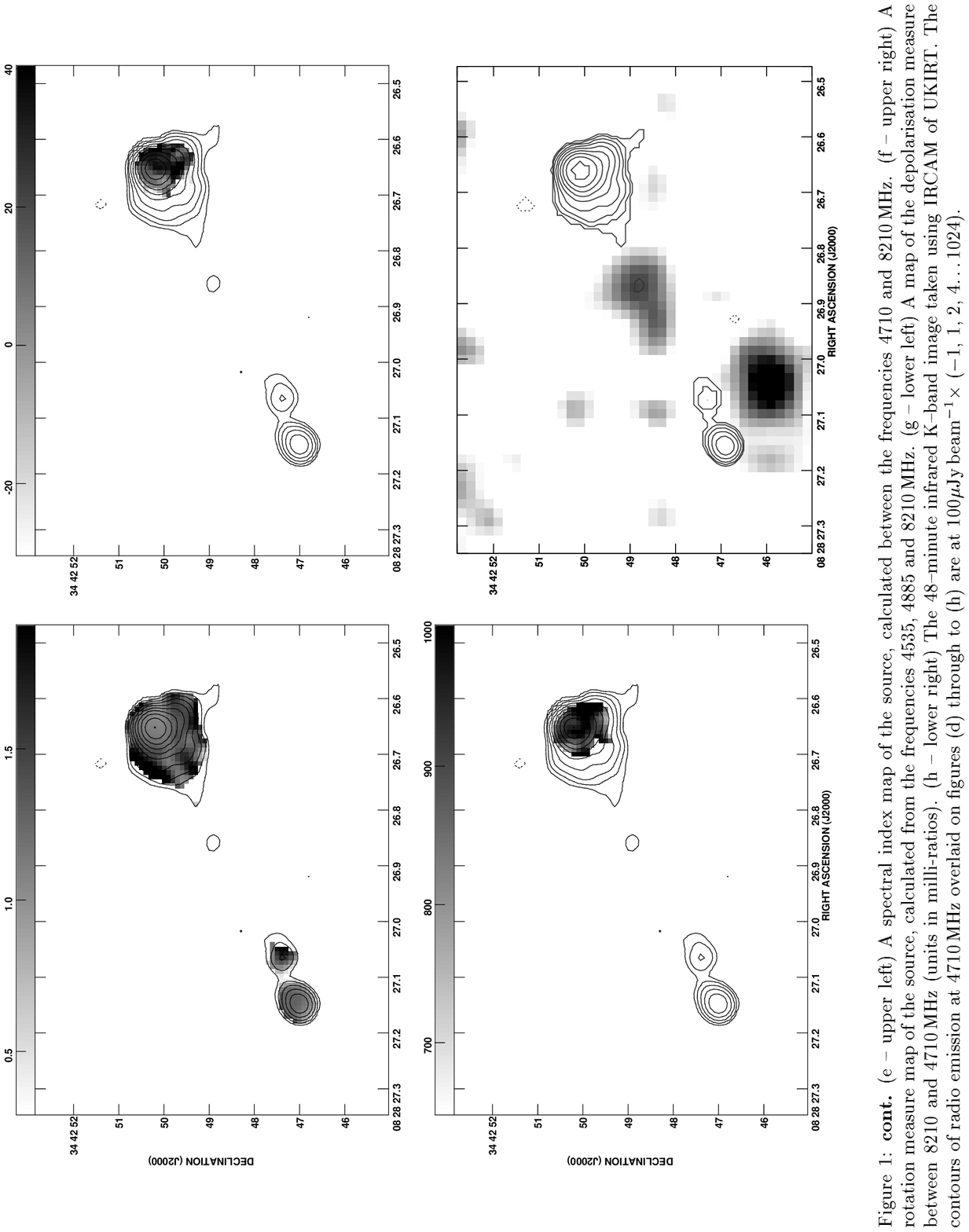,clip=,width=\textwidth}
}
\end{figure*}

\begin{figure*}
\centerline{
\psfig{figure=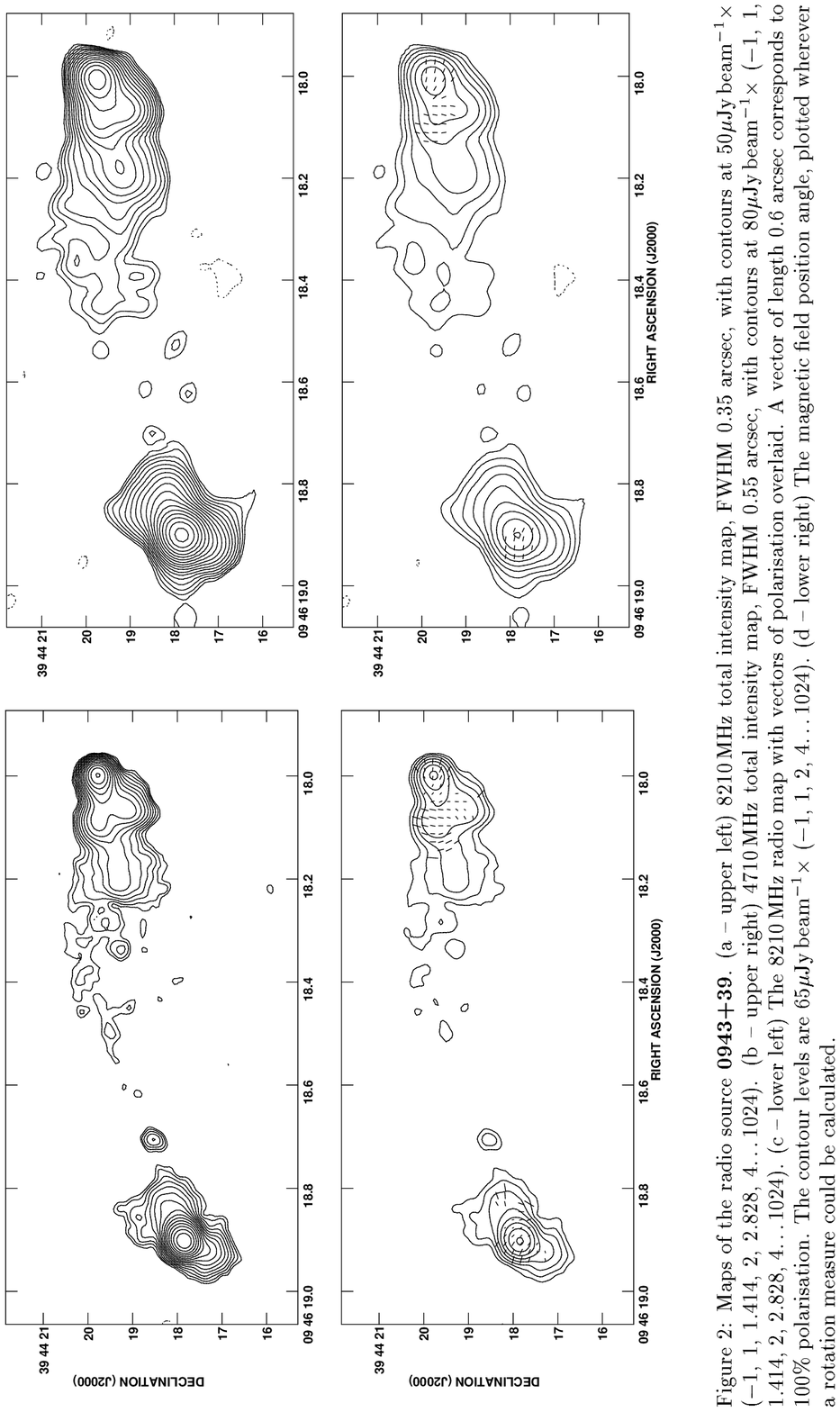,clip=,width=14.2cm}
}
\end{figure*}

\begin{figure*}
\centerline{
\psfig{figure=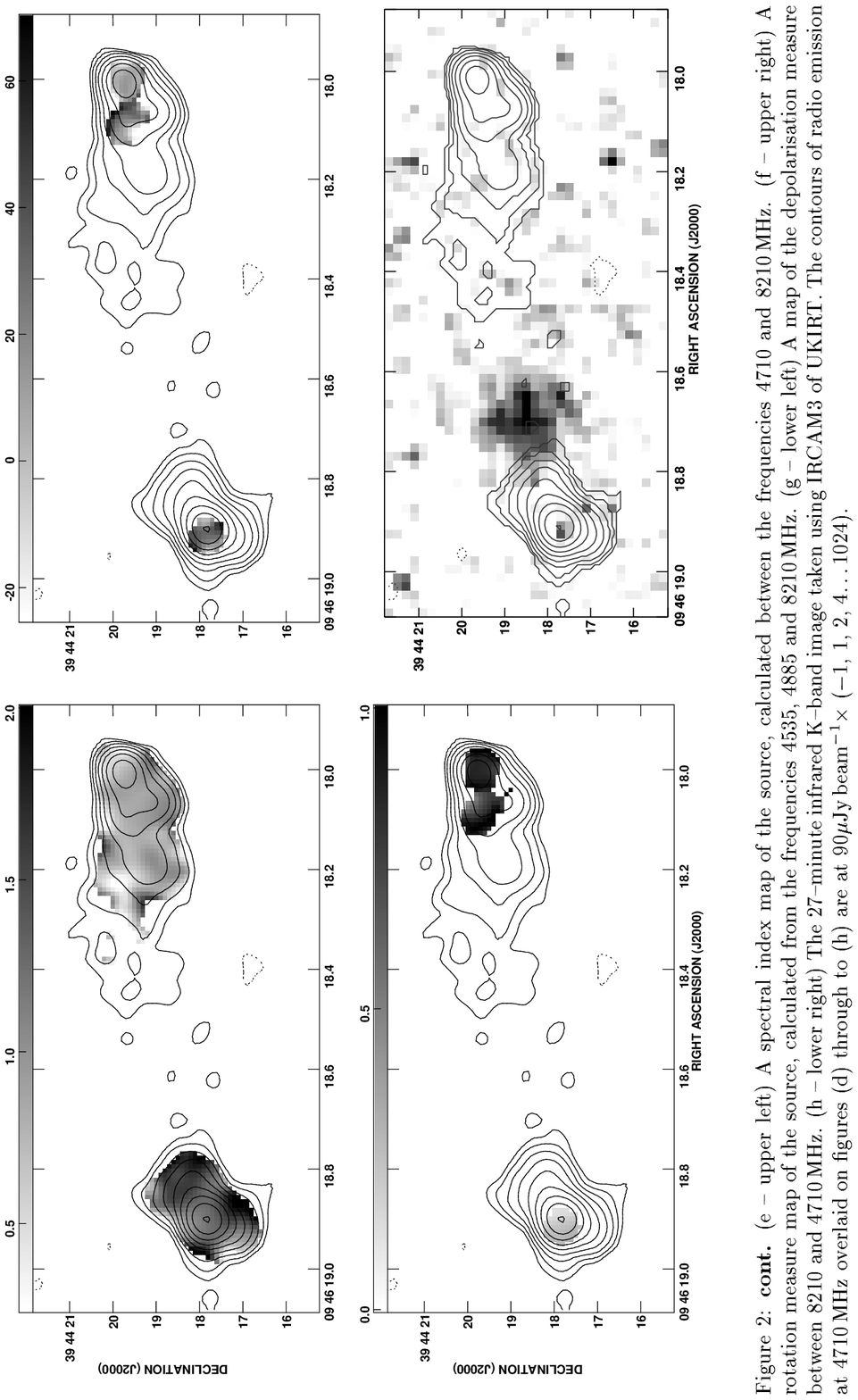,clip=,width=14.7cm}
}
\end{figure*}

\begin{figure*}
\centerline{
\psfig{figure=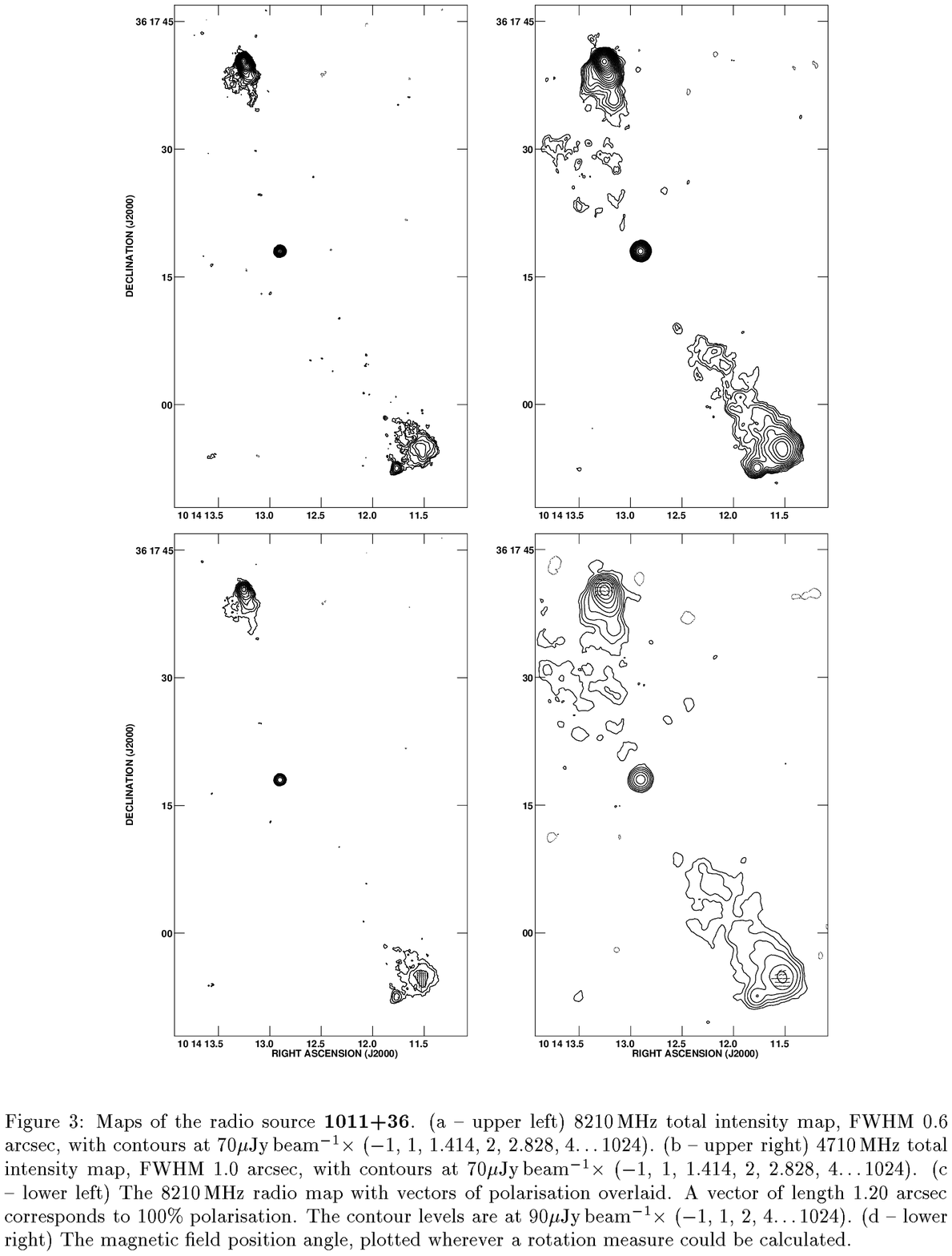,clip=,width=\textwidth}
}
\end{figure*}

\begin{figure*}
\centerline{
\psfig{figure=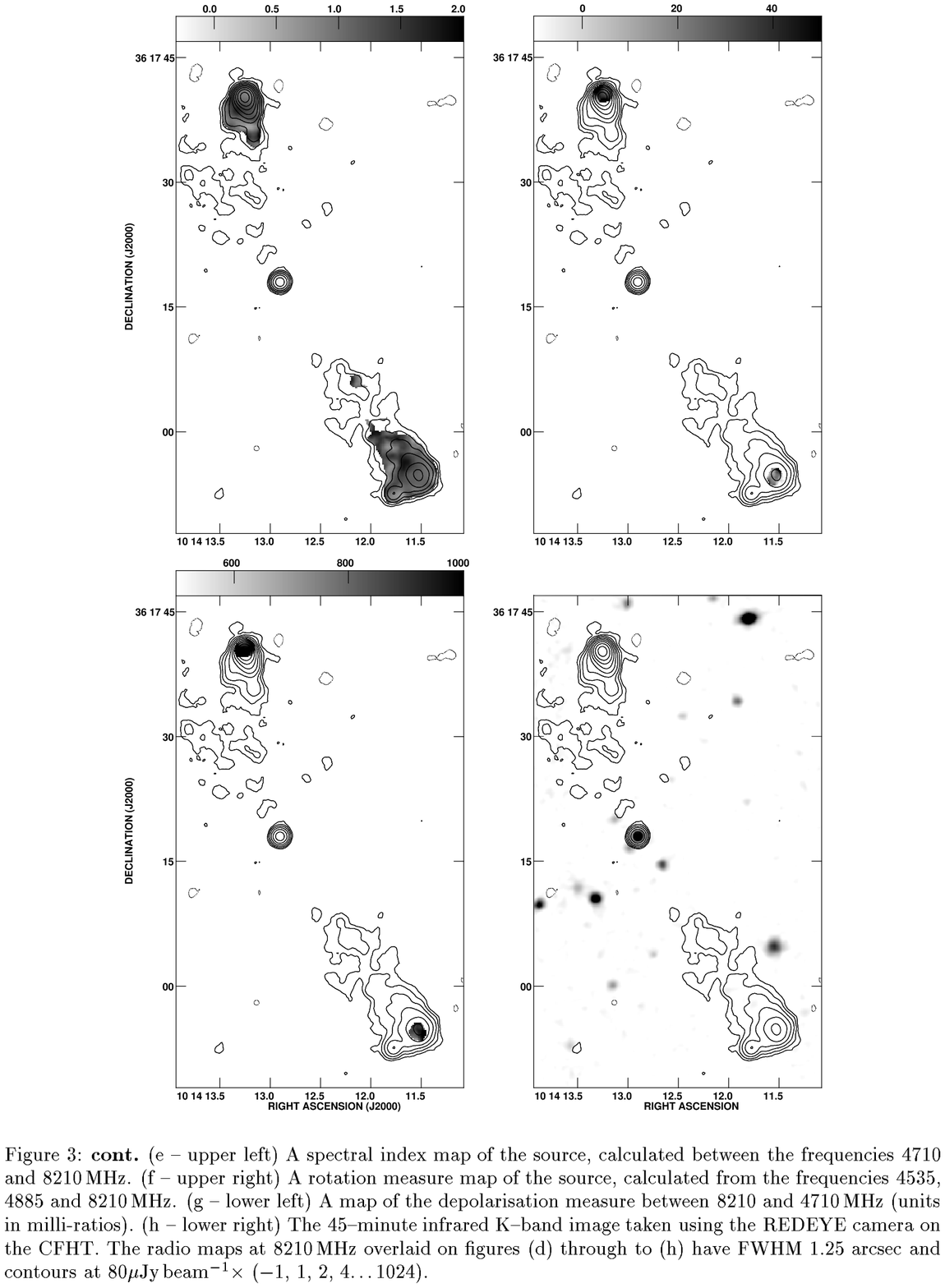,clip=,width=\textwidth}
}
\end{figure*}

\begin{figure*}
\centerline{
\psfig{figure=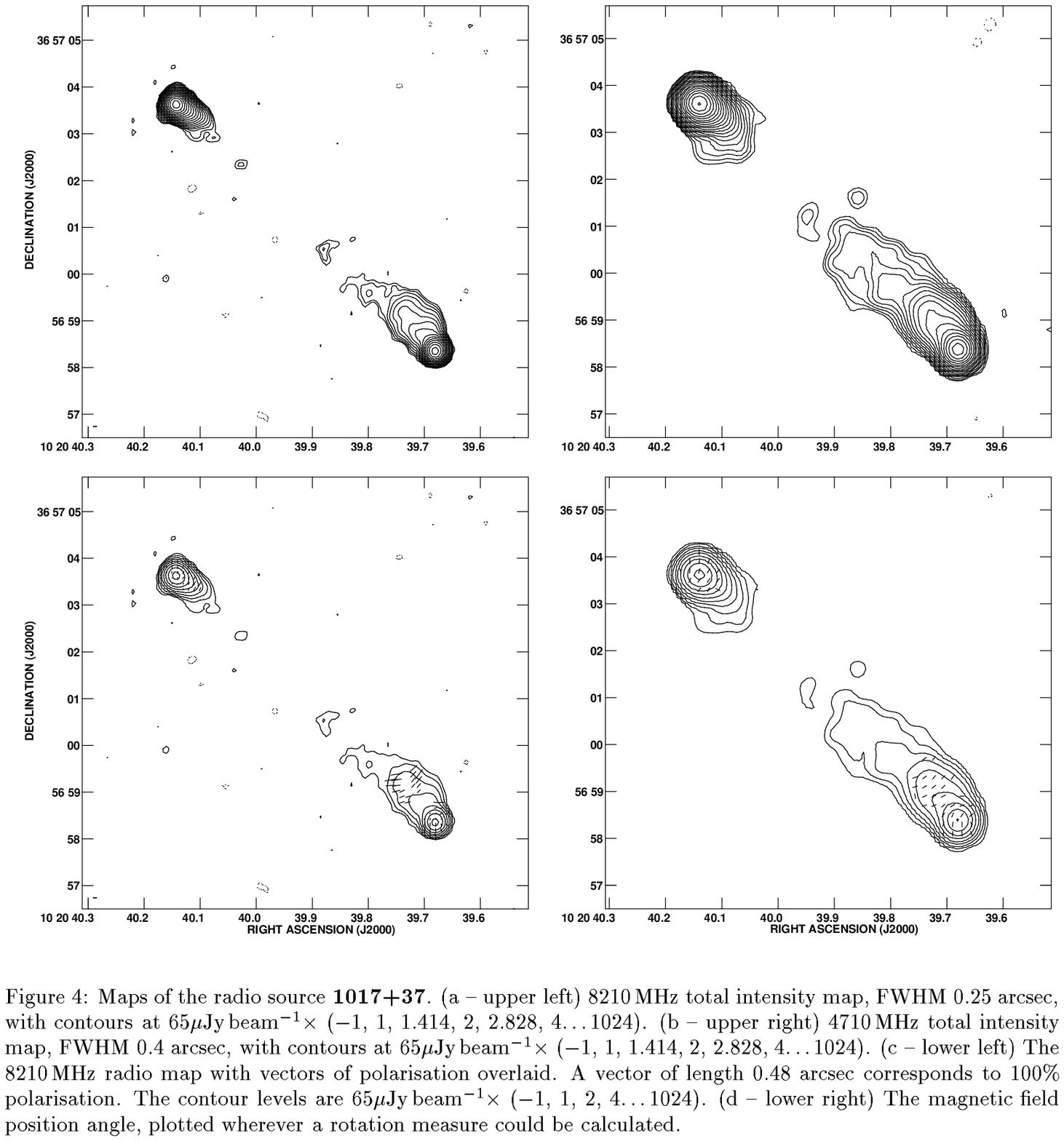,clip=,width=\textwidth}
}								
\end{figure*}							
								
\begin{figure*}							
\centerline{							
\psfig{figure=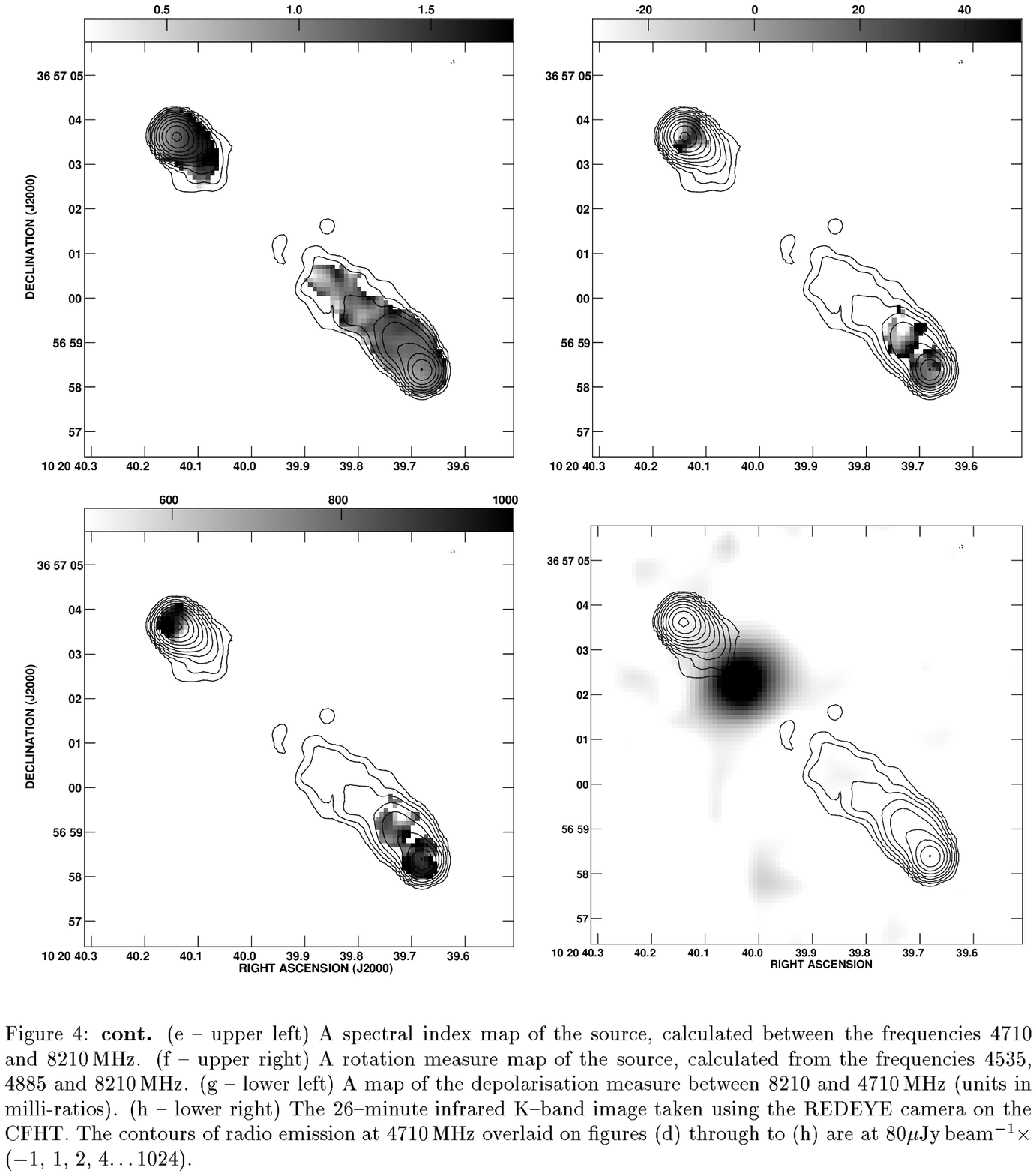,clip=,width=\textwidth}
}								
\end{figure*}							
								
\begin{figure*}							
\centerline{							
\psfig{figure=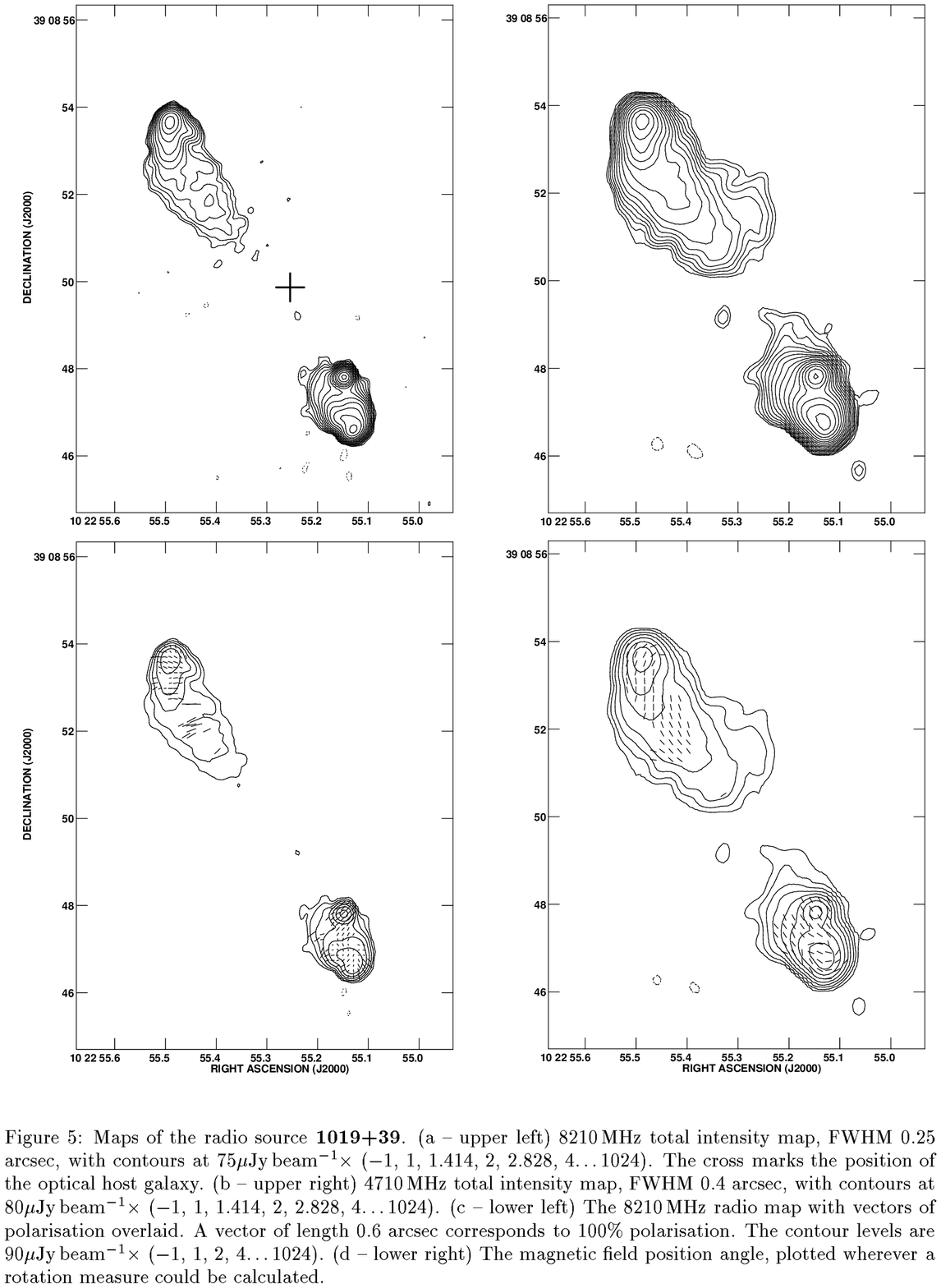,clip=,width=17cm}
}								
\end{figure*}							
								
\begin{figure*}							
\centerline{							
\psfig{figure=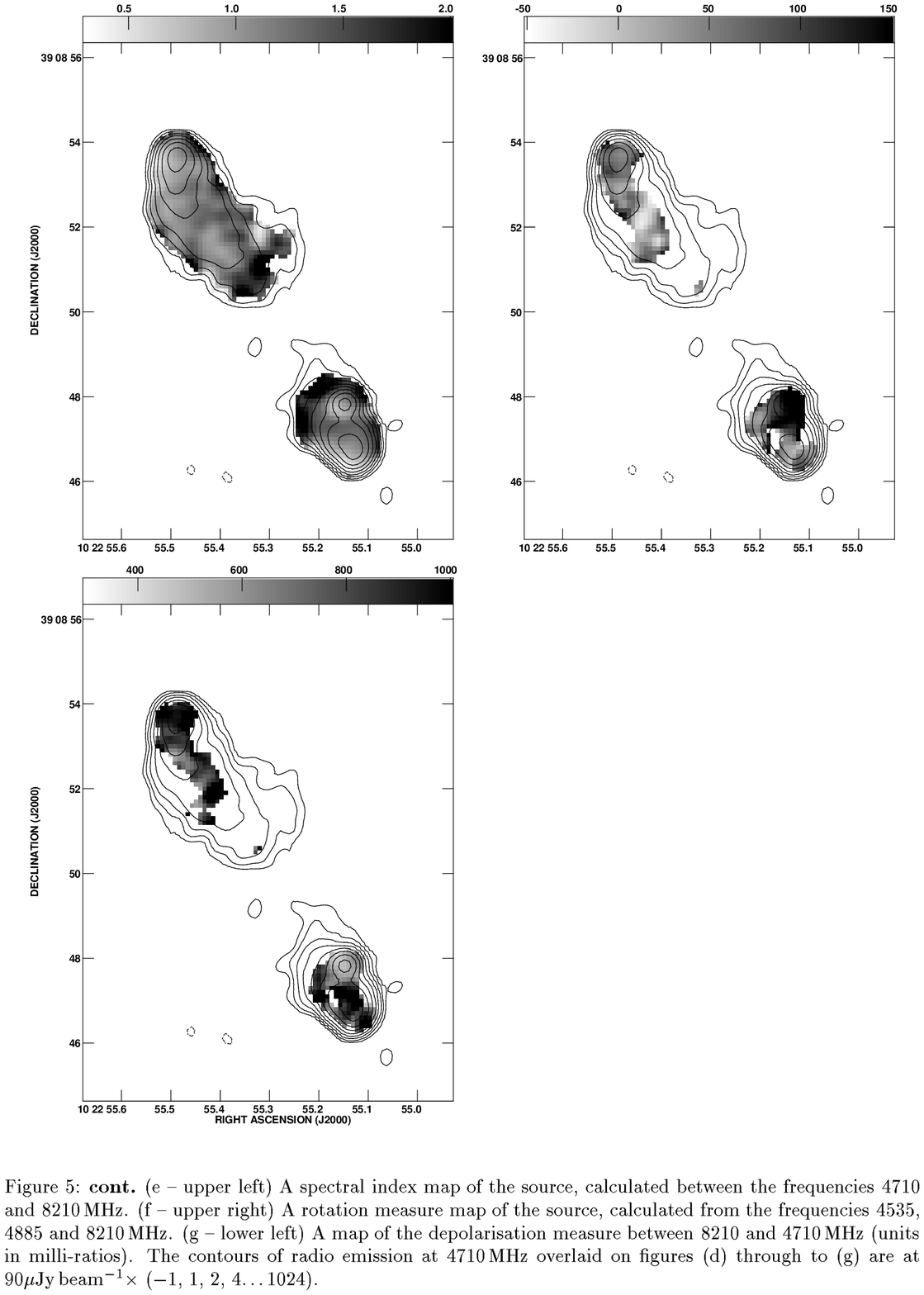,clip=,width=17cm}
}
\end{figure*}

\begin{figure*}
\centerline{
\psfig{figure=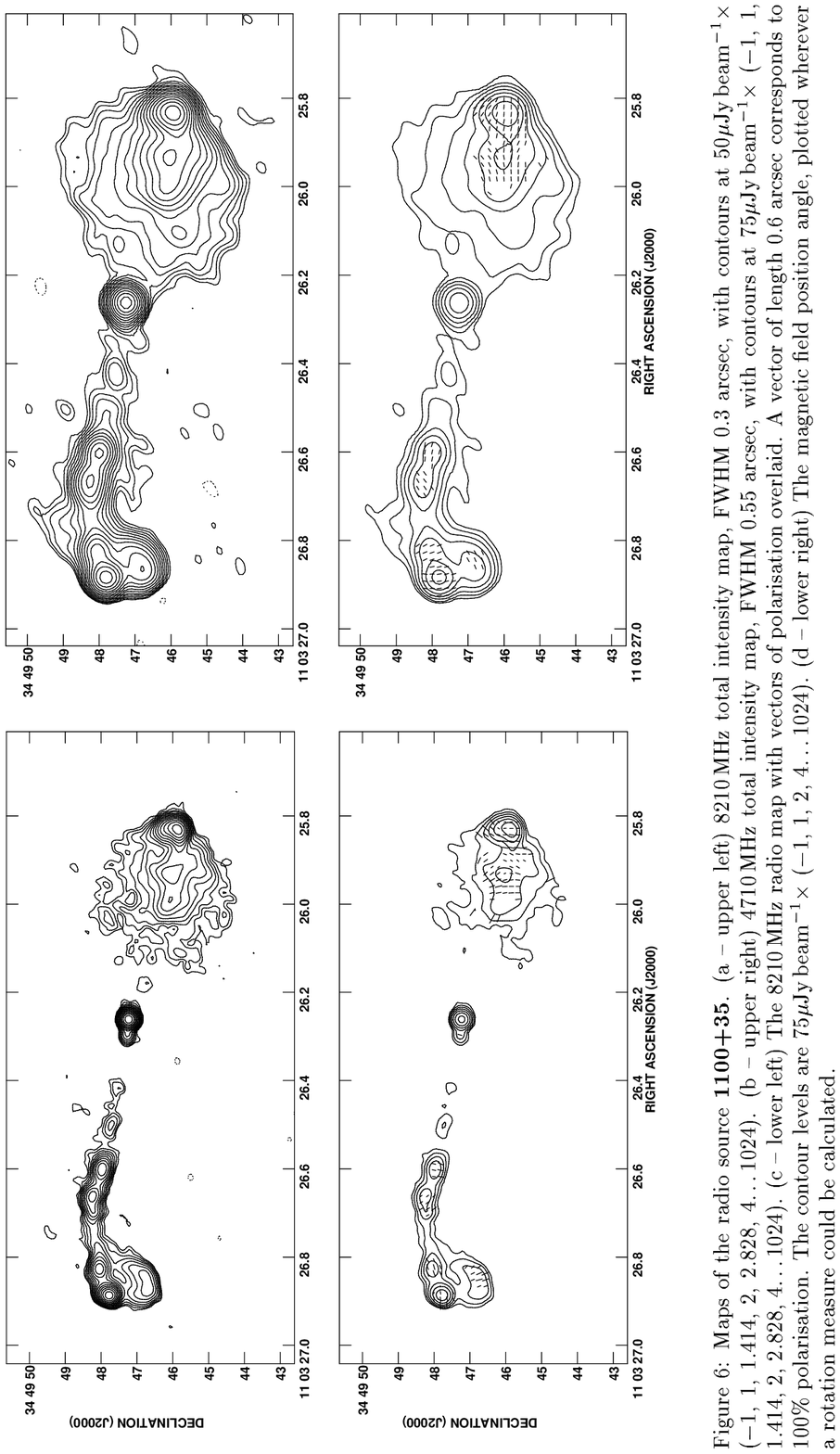,clip=,width=13.5cm}
}
\end{figure*}

\begin{figure*}
\centerline{
\psfig{figure=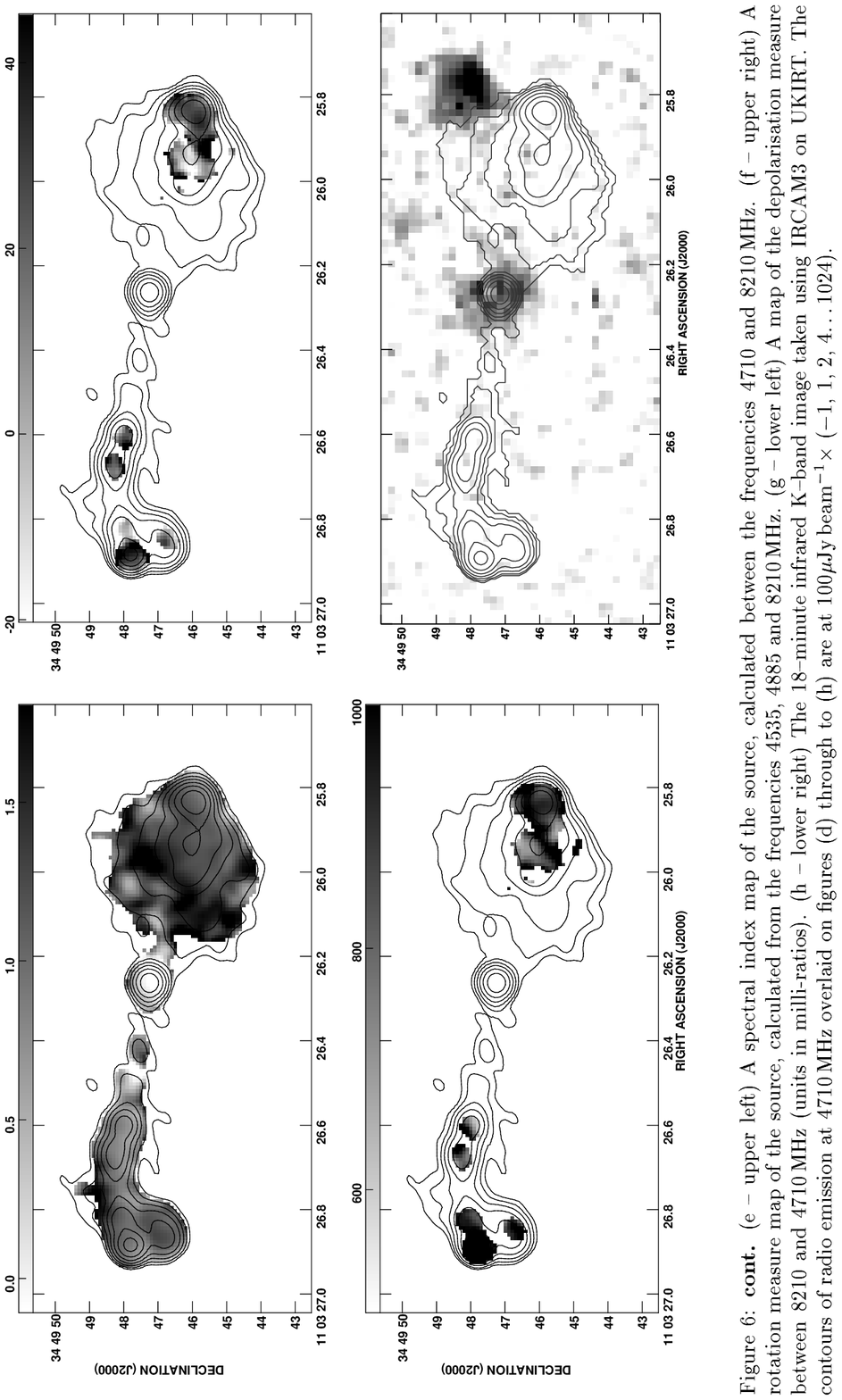,clip=,width=14.3cm}
}
\end{figure*}

\begin{figure*}
\centerline{
\psfig{figure=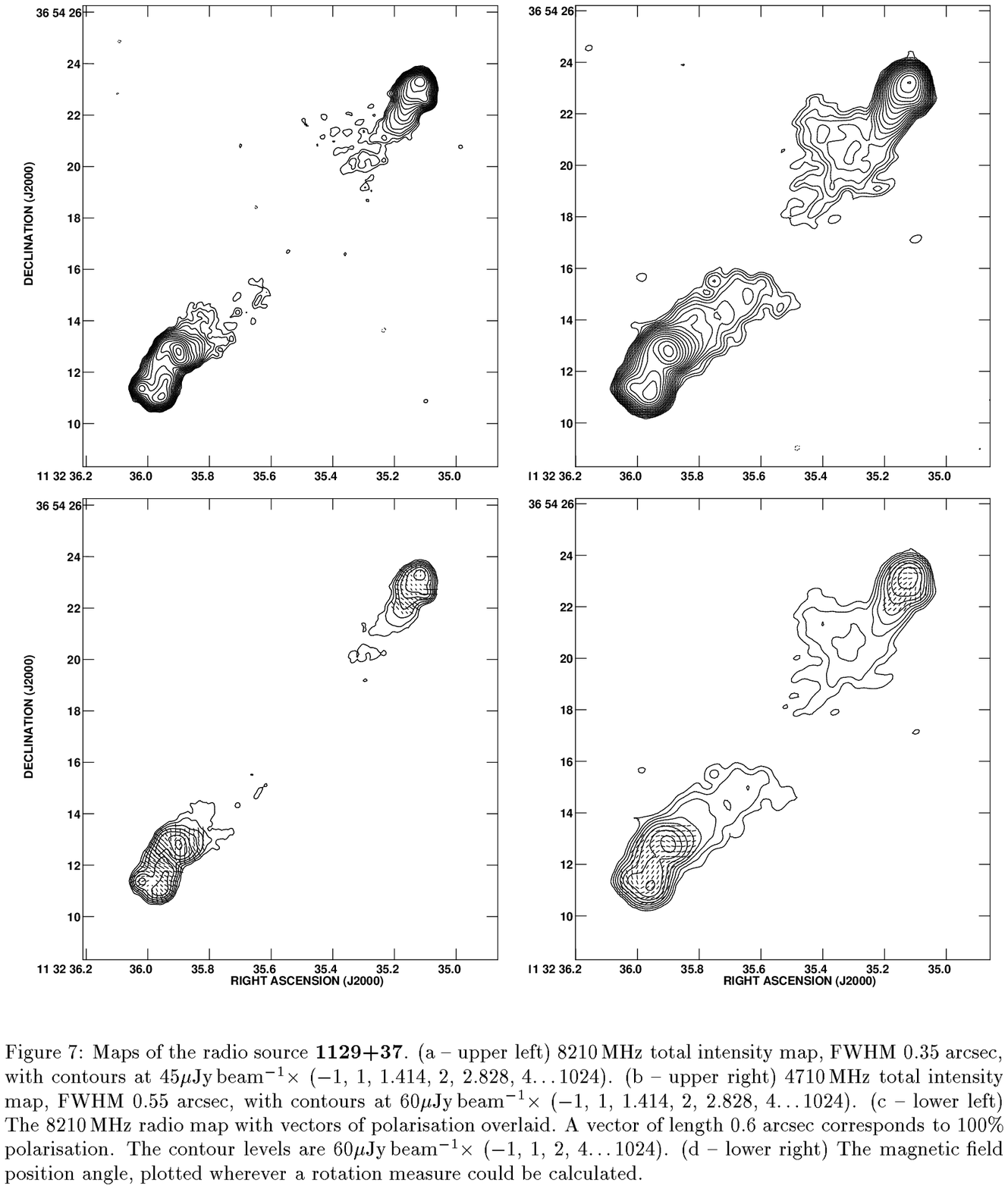,clip=,width=\textwidth}
}								
\end{figure*}							
								
\begin{figure*}							
\centerline{							
\psfig{figure=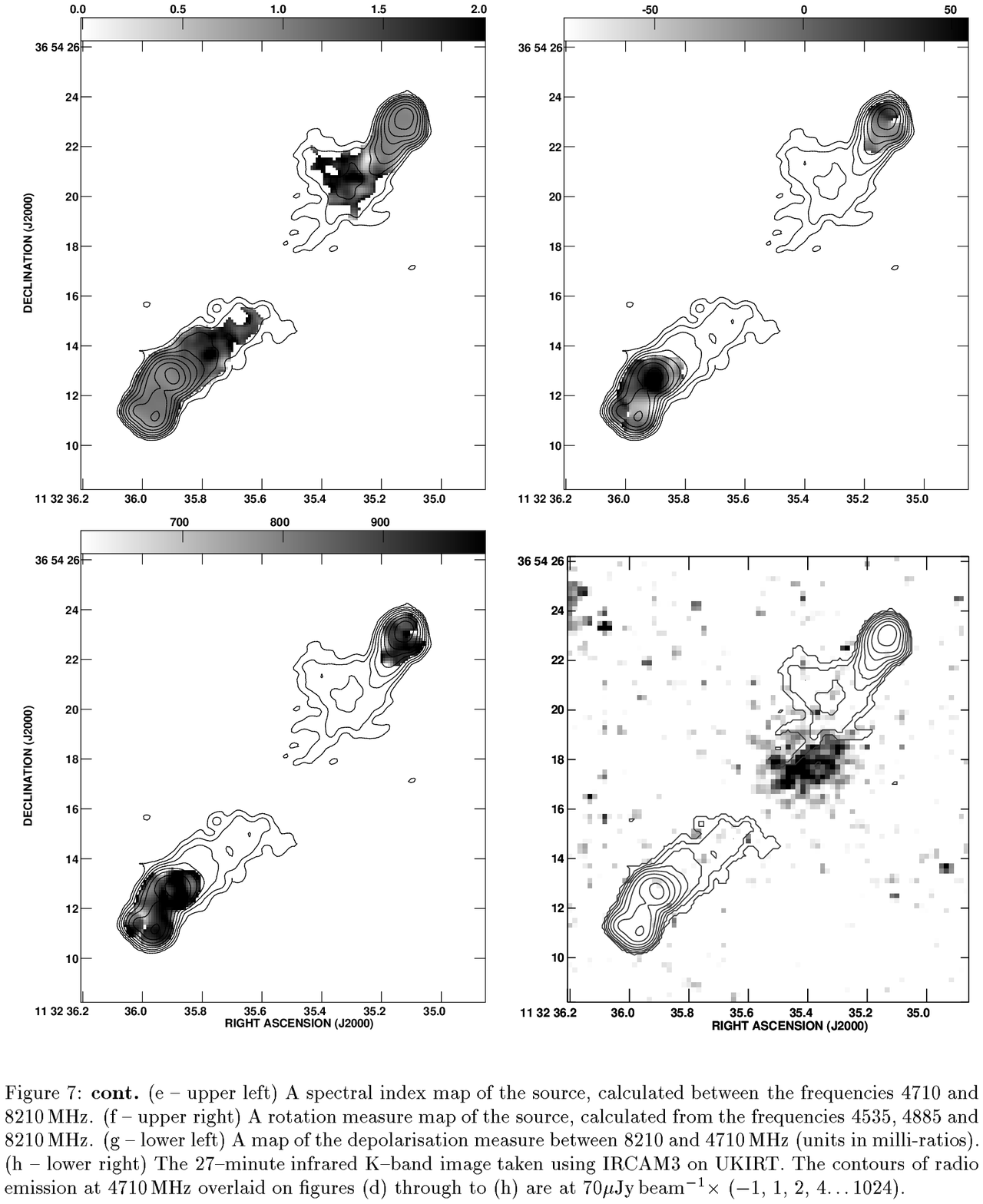,clip=,width=\textwidth}
}
\end{figure*}

${\it 1129+37:}$ No radio core is detected for this object in the current
observations, although the host galaxy identification by Allington--Smith
\shortcite{all82b} is secure and the galaxy has a redshift of 1.06
\cite{raw98}. The south--eastern lobe of the radio
source contains three hotspots and is strongly polarised with a
well-defined magnetic field structure. The hotspot closest to the nucleus
corresponds to a region of significantly higher rotation measure than the
two more distant hotspots, by as much as a few hundred rad\,m$^{-2}$ in
the source rest-frame. The north--western lobe is more regular and has a
lower polarisation.

\bigskip

${\it 1204+35:}$ This source contains a strong radio core (see also
Law--Green \etal\ 1995)\nocite{law95} which, although it has a relatively
steep radio spectrum, is co-incident with the optical
identification of Allington--Smith \shortcite{all82b}. Rawlings \etal\
\shortcite{raw98} have determined the redshift of the host galaxy to be
$z=1.37$. Both radio lobes contain double hotspots and show moderate 
depolarisation.
\bigskip

\begin{figure*}
\centerline{
\psfig{figure=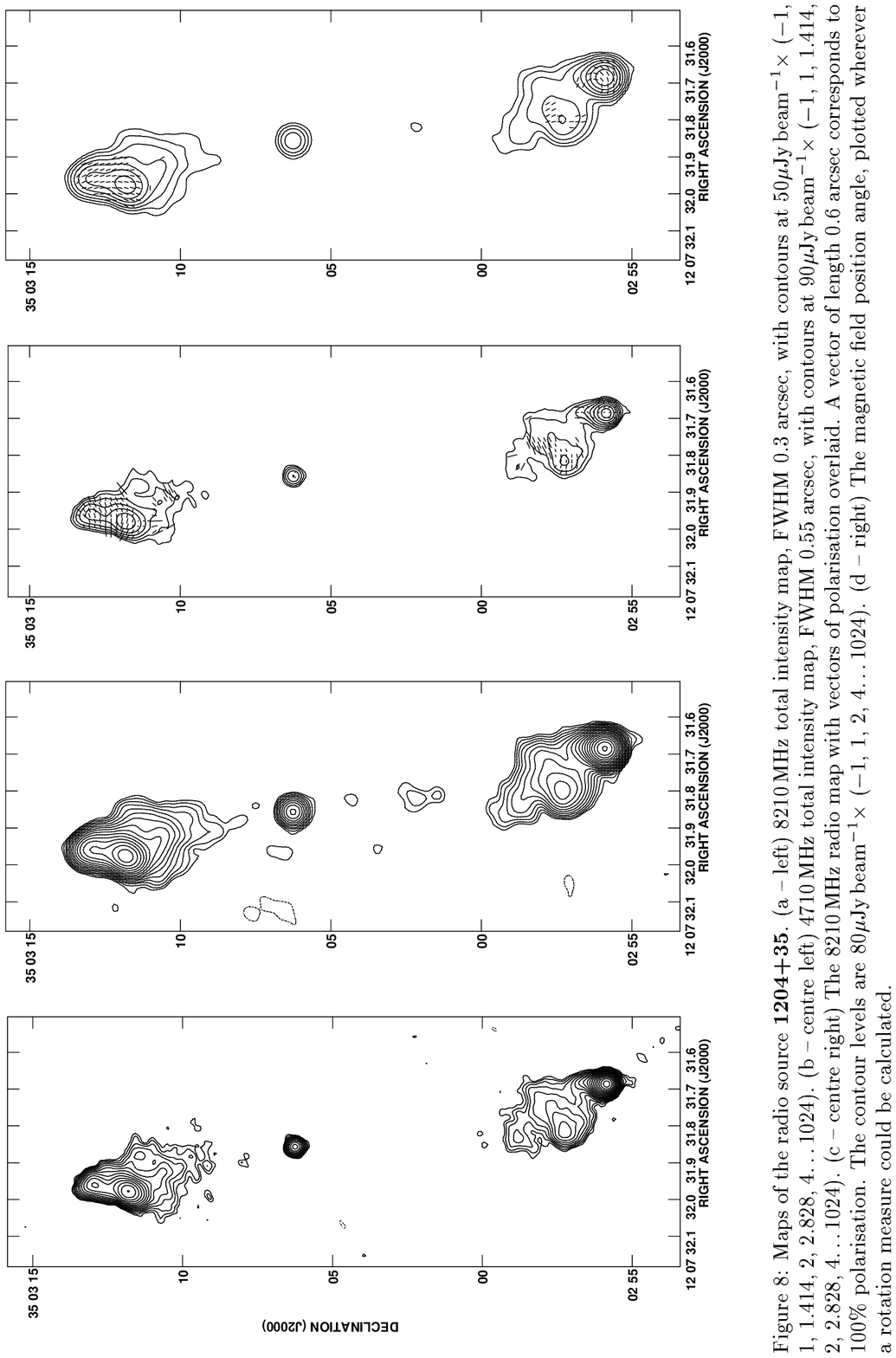,clip=,width=15cm}
}
\end{figure*}

\begin{figure*}
\centerline{
\psfig{figure=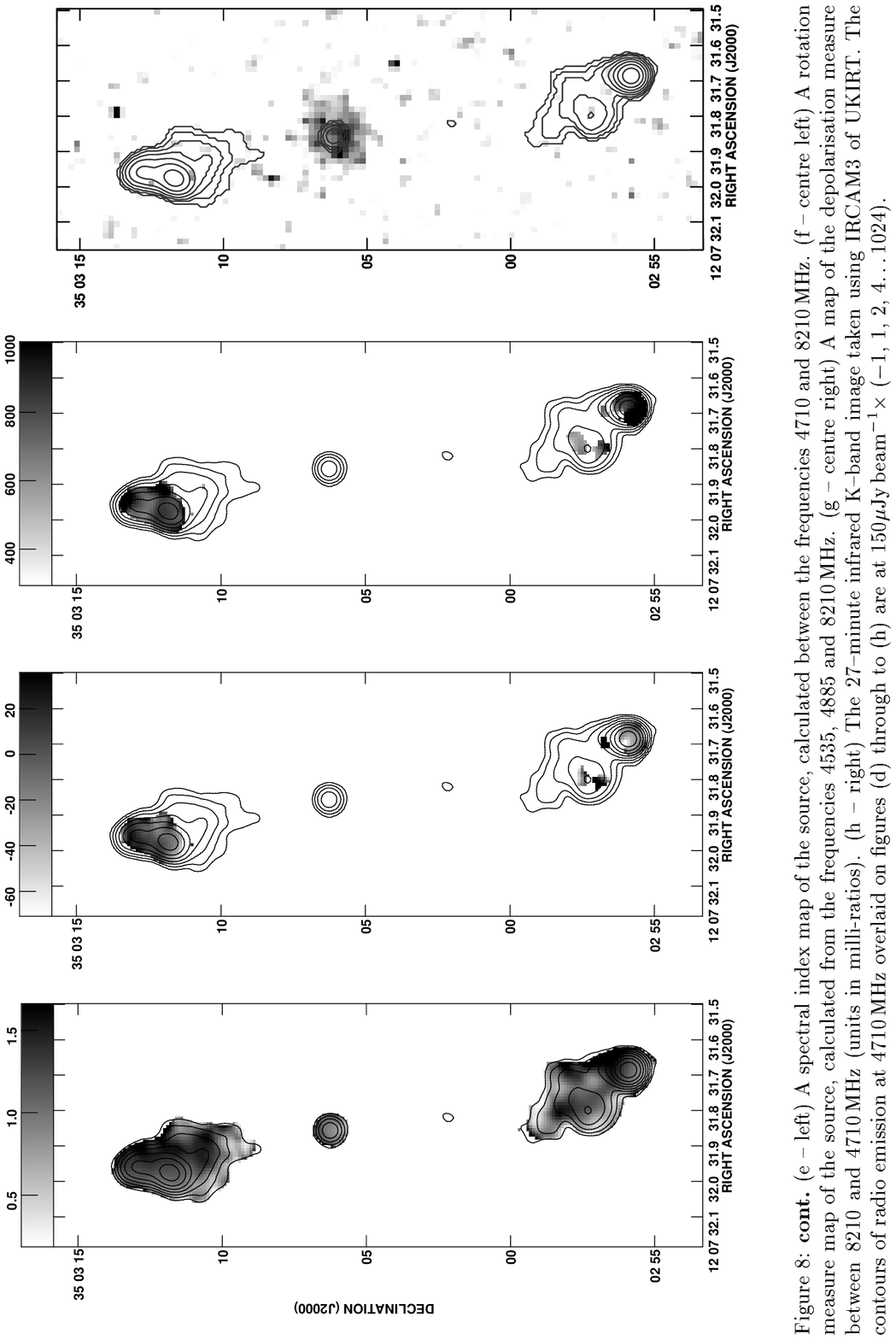,clip=,width=16cm}
}
\end{figure*}

${\it 1217+36:}$ This source has a very peculiar radio morphology (see
also Naundorf \etal\ 1992; Law--Green \etal\ 1995).\nocite{nau92,law95} A
central compact double, 0.7 arcsec in extent, is surrounded by an halo of
emission extending more than 4 arcsecs. The spectral index of this radio
halo steepens with distance out from the central object; if we had not
restricted the spectral index map to regions where flux was detected at
greater than the 5-sigma level in both maps, then it could have been seen
to steepen to values of 2.5 or more in the outer regions of the halo. The
north--eastern of the two compact emission knots has a relatively flat
spectral index and probably includes the radio core, but the brightness of
this region, its resolved size, and the detection of polarised emission
from it suggests that this knot does not contain solely the radio
core. Also interesting is that the region of flattest spectral index
appears to be offset by about 0.5 arcsec east from the peak of the radio
flux density.

\bigskip
\begin{figure*}
\centerline{
\psfig{figure=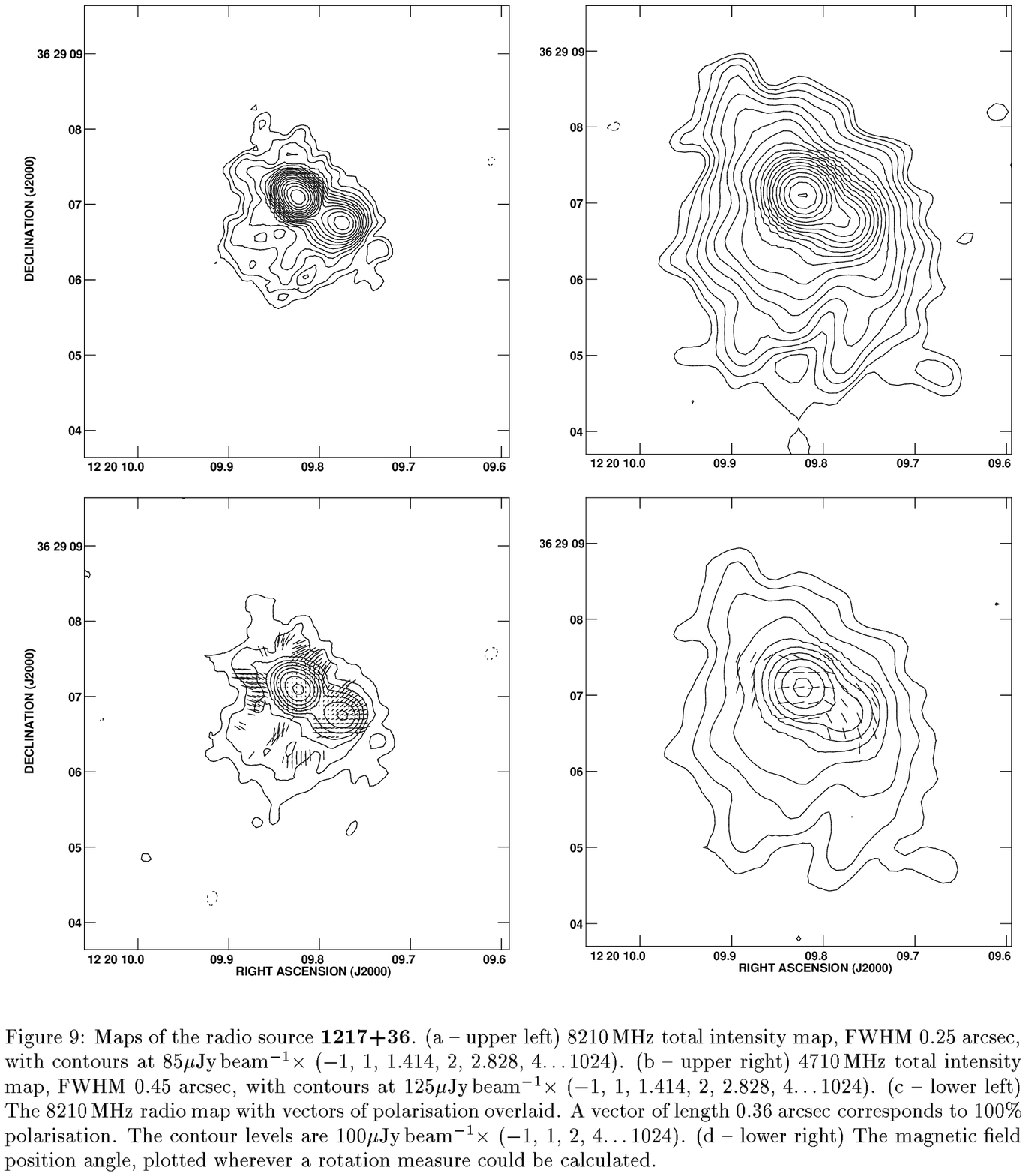,clip=,width=\textwidth}
}								
\end{figure*}							
								
\begin{figure*}							
\centerline{							
\psfig{figure=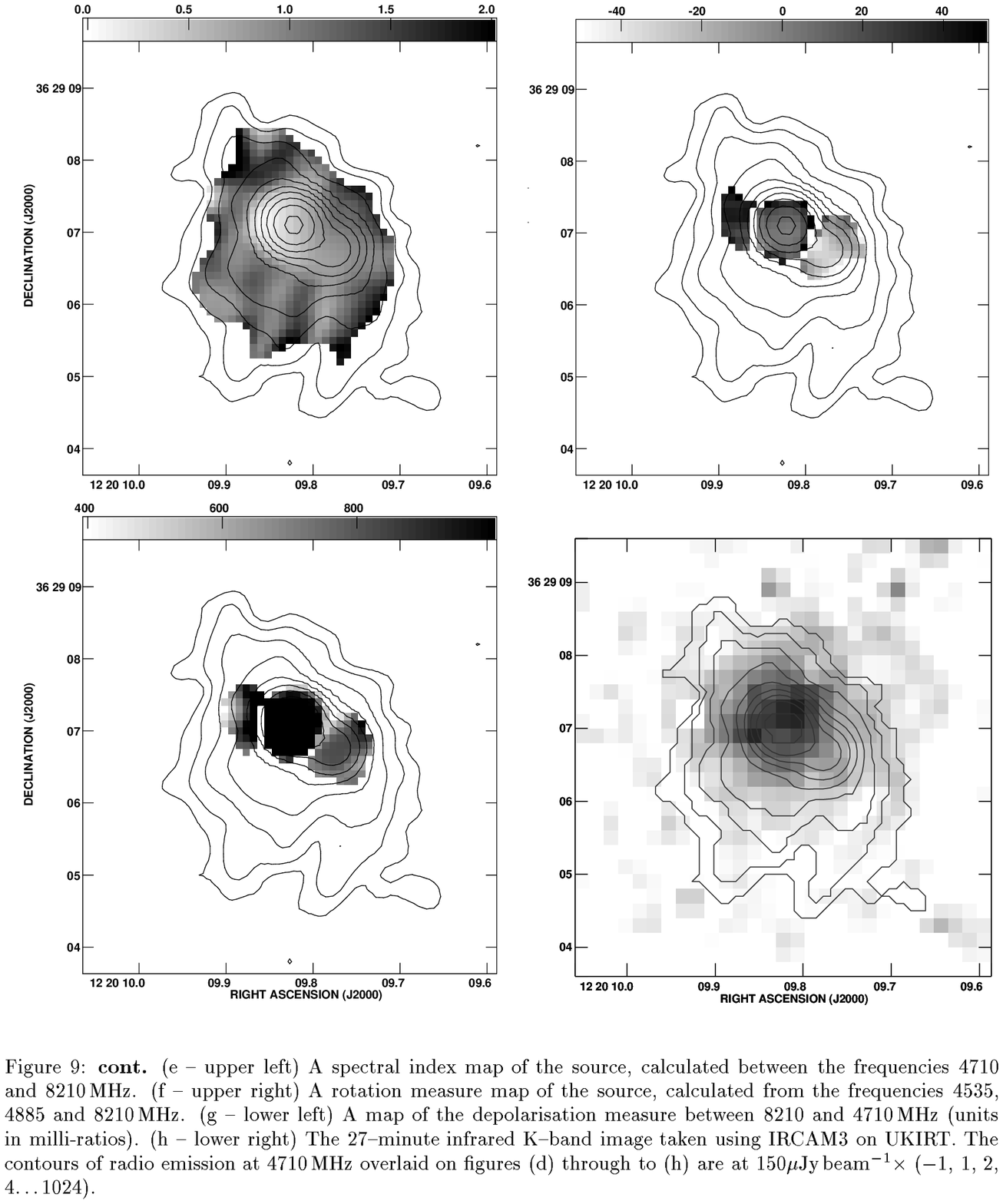,clip=,width=\textwidth}
}								
\end{figure*}							

\begin{figure*}							
\centerline{							
\psfig{figure=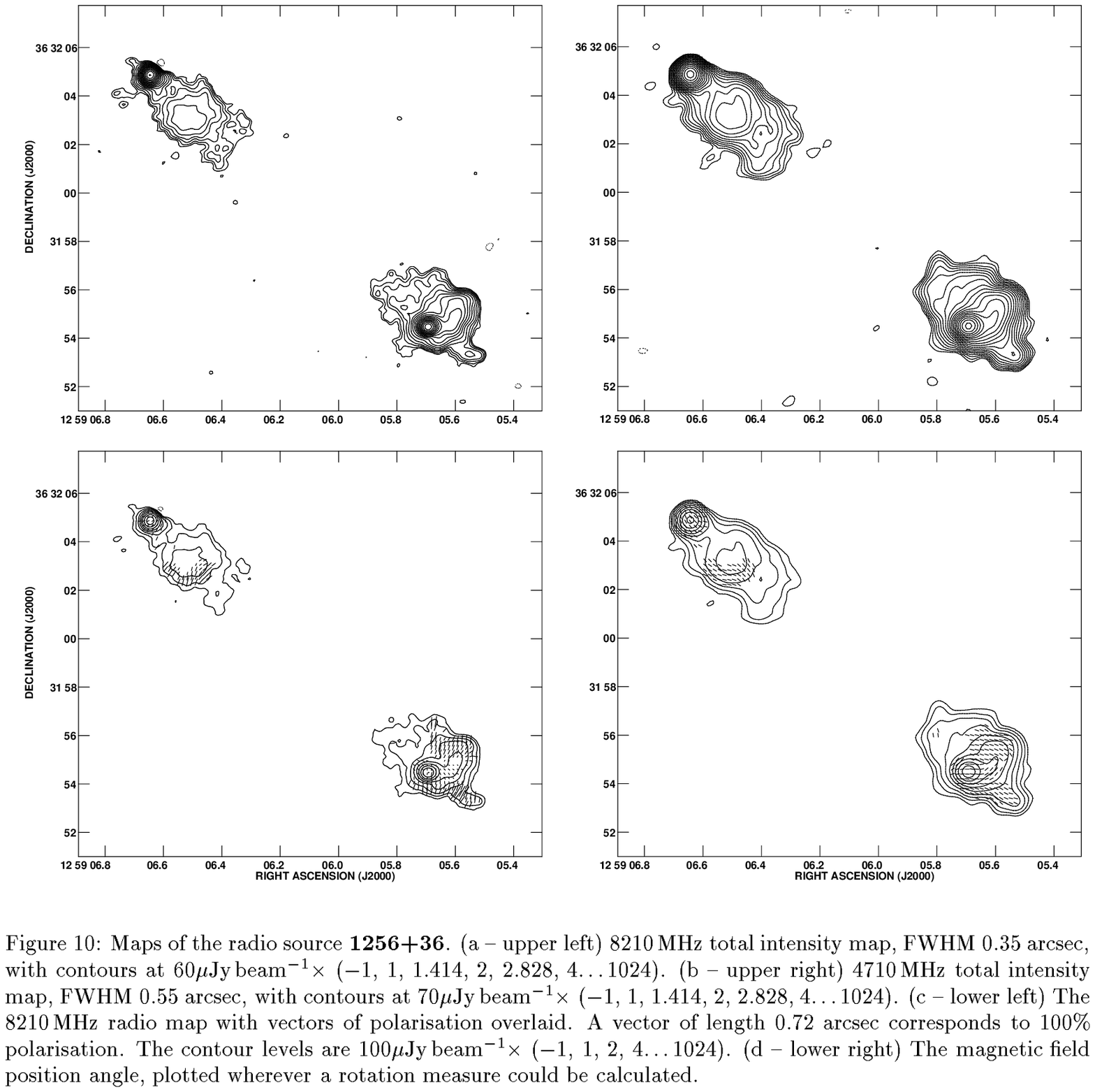,clip=,width=\textwidth}
}								
\end{figure*}							
								
\begin{figure*}							
\centerline{							
\psfig{figure=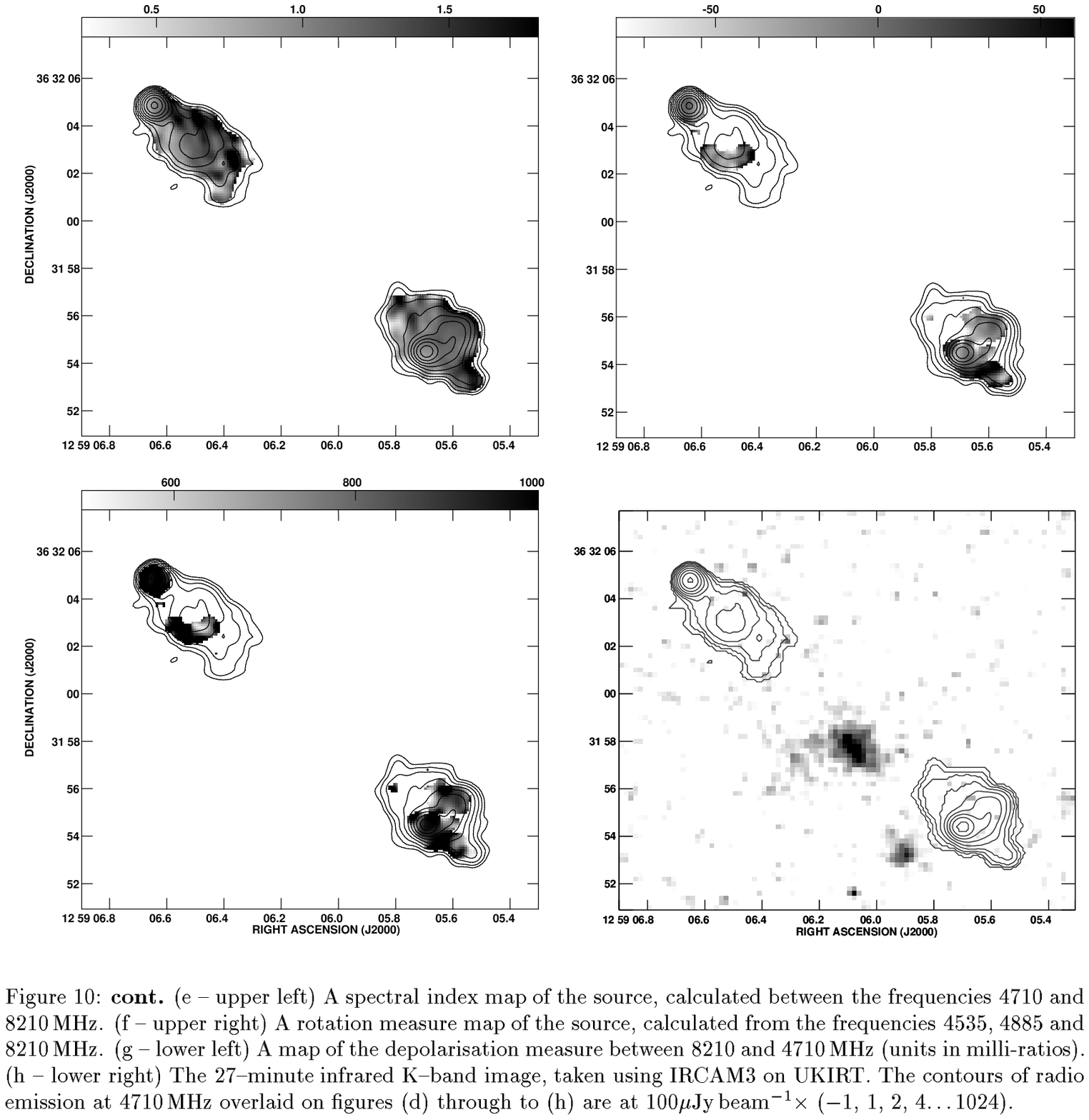,clip=,width=\textwidth}
}								
\end{figure*}							
								
\begin{figure*}							
\centerline{							
\psfig{figure=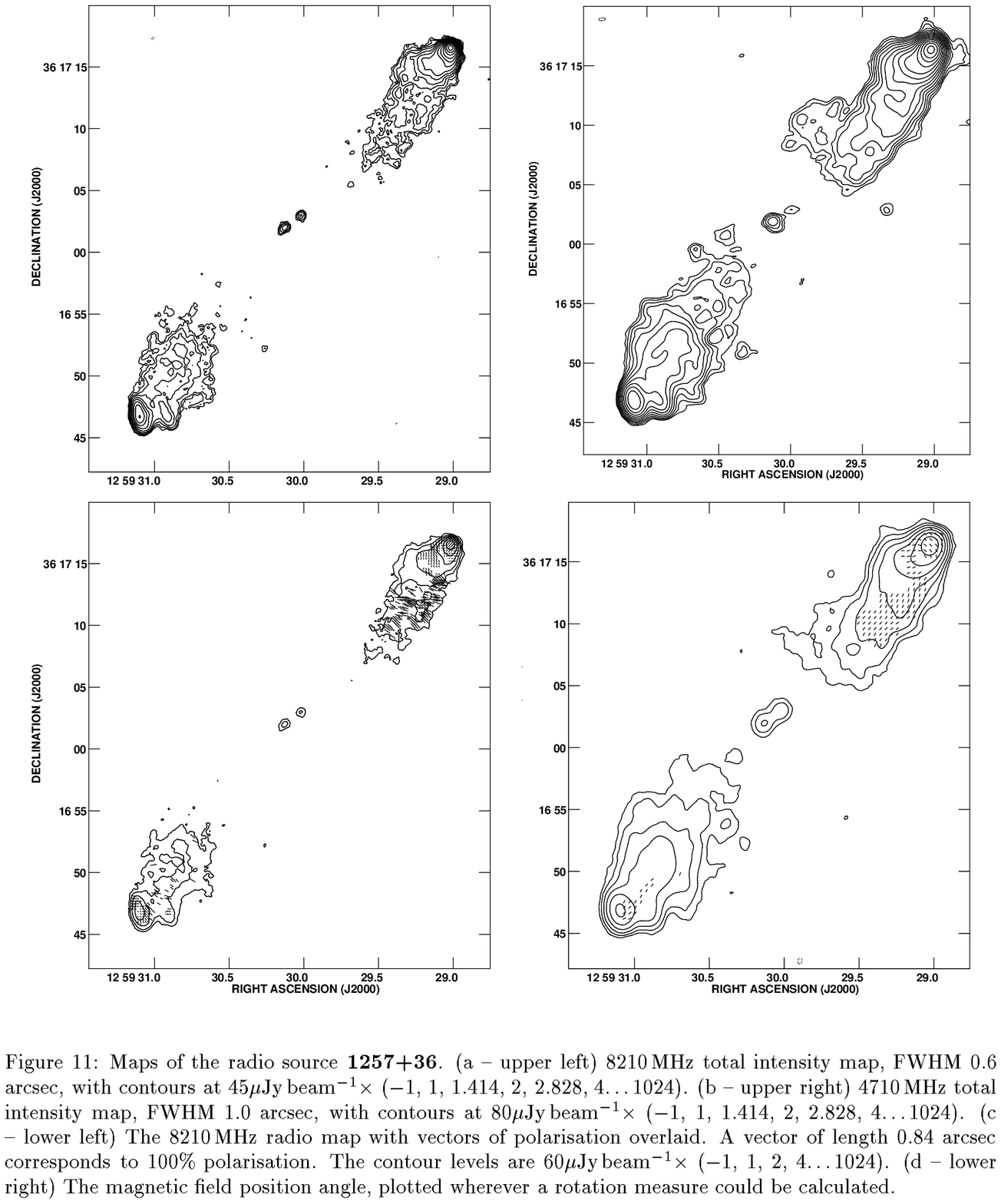,clip=,width=\textwidth}
}								
\end{figure*}							
								
\begin{figure*}							
\centerline{							
\psfig{figure=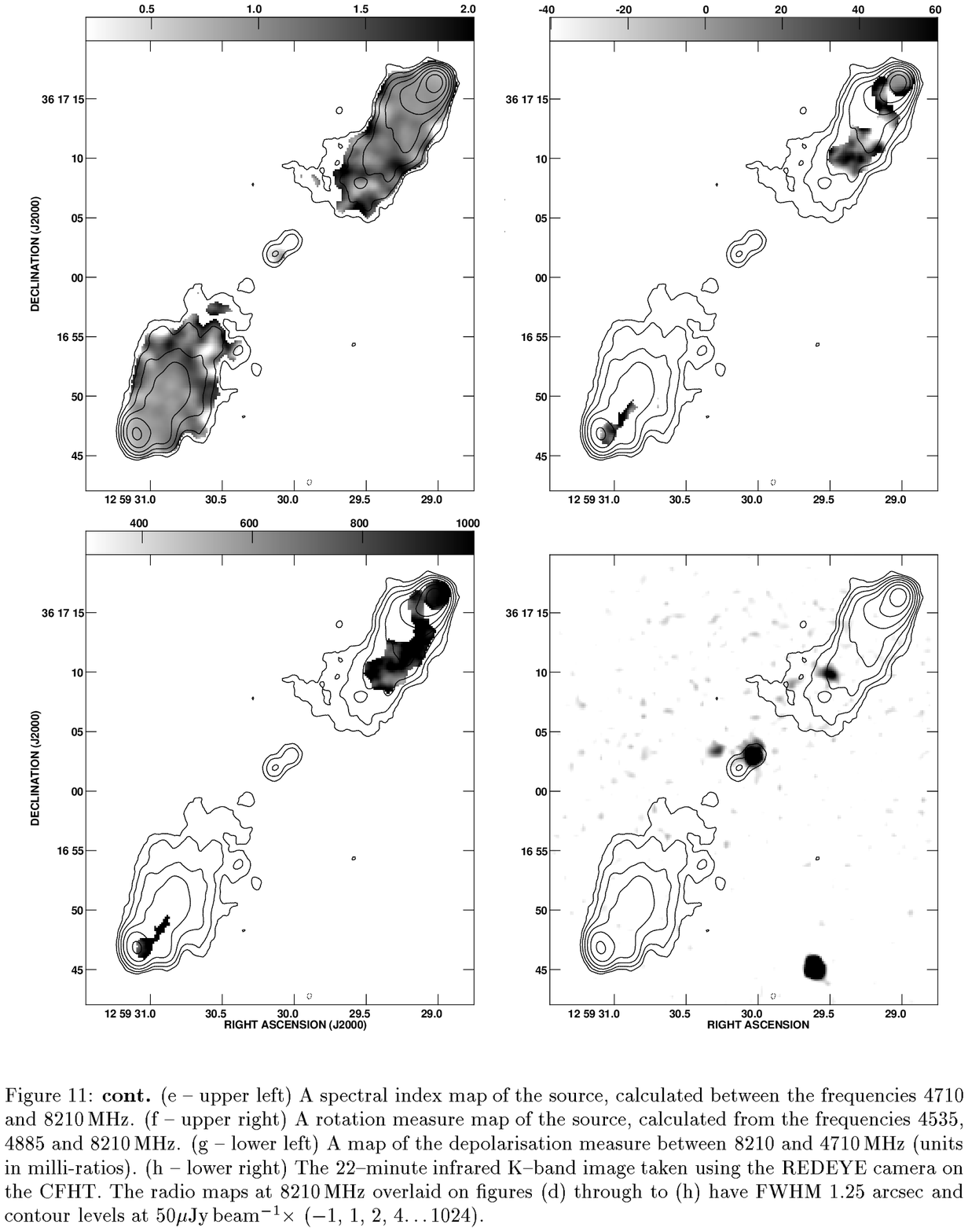,clip=,width=\textwidth}
}
\end{figure*}

The emission from the central region of this source shows no
depolarisation, which is surprising given that this should lie well within
the gas of the host galaxy. The diffuse emission halo is more strongly
depolarised. There is also a steep gradient in the rotation measure across
the source.

Ben\'itez \etal\ \shortcite{ben95} suggested that this galaxy was
stellar--like on their R--band image. They therefore proposed that it
should perhaps be reclassified as a quasar and that the redshift of 1.2
estimated from its K--magnitude \cite{lil89} would be too high. A spectrum
by Rawlings \etal\ \shortcite{raw98} has provided a redshift of $z=1.09$,
close to Lilly's estimate (although based upon only one emission
line). Neither this spectrum nor the K--band image suggest that this
source is a quasar.

\bigskip

${\it 1256+36:}$ This radio source was identified by Lilly
\shortcite{lil89} with a galaxy whose redshift has been determined as
$z=1.07$ \cite{raw98}. Although no radio core is detected on our radio
maps, the identification is secure, lies directly between the two radio
lobes, and appears to show an elongation along the radio axis. Lilly
\shortcite{lil89} detected five other galaxies within a few arcseconds of
the host, possibly forming the core of a rich group or cluster; only two
of these are detected in our (albeit slightly shallower) K--band image.
The hotspot in the south--western radio lobe is withdrawn from the leading
edge of the emission (see also Law--Green \etal\ 1995).\nocite{law95} The
two lobes show only small variations between their spectral indices,
depolarisation and rotation measures.

\bigskip

${\it 1257+36:}$ Two core candidates, separated by only 1.5 arcseconds,
lie towards the centre of this radio source and close to the host galaxy
identified by Eales \etal\ \shortcite{eal97} at redshift $z=1.00$
\cite{raw98}. The south--eastern core candidate is the brighter; the
north--western has the flatter spectral index. With the current data it is
impossible to unambiguously determine which is the core of the radio
galaxy. We have aligned the infrared image by assuming, arbitrarily, that
the north--western component is the core.

The second largest source in our sample, this radio source extends nearly
40 arcseconds. The south--eastern lobe contains a double hotspot and shows
very little polarised emission. The north-western lobe is more regular
both in its structure and polarisation properties. Note that, as discussed
at the beginning of this Section, the slightly speckled nature of the
spectral index map close to the extremities of both lobes is likely to be
an artefact caused by the inability of the \clean ing process to
accurately represent smooth extended low--surface brightness emission.
\bigskip
\addtocounter{figure}{11}

\section{Discussion}
\label{discuss}

A number of features stand out in the radio data presented, and here we
discuss these features and compare them with our sample of 3CR radio
galaxies at redshift $z \sim 1$ \cite{bes97c}, and with a sample of low
redshift radio galaxies.

\begin{figure}
\centerline{
\psfig{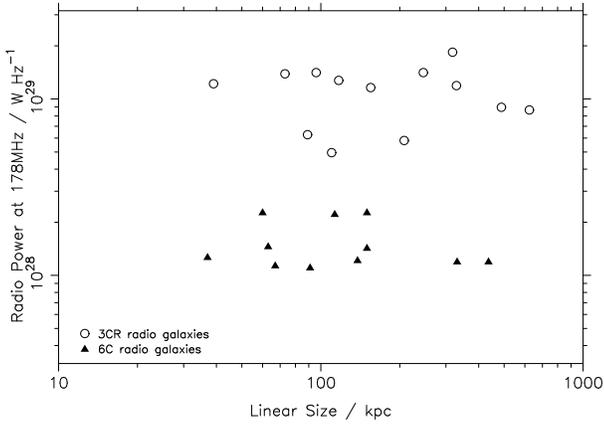}
}
\caption{\label{pddiag} The radio power ($P$) versus linear size ($D$) diagram
from the 6C radio galaxies presented here and the 3CR radio galaxies from
the sample of Best \etal\ (1997) in the overlapping redshift range.}
\end{figure}

A comparison of the low--frequency radio power ($P$) {\it vs} linear size
($D$) diagram for the 6C radio galaxies and those in the 3CR sample within
the overlapping redshift range $0.85 \le z \le 1.5$ is shown in
Figure~\ref{pddiag}, clearly demonstrating the lower radio power of the 6C
radio galaxies. The average radio size of the 6C sources ($D_{\rm mean} =
137 \pm 35$\,kpc; $D_{\rm med} = 102$\,kpc, after re-including 1123+34
excluded from the sample by an angular size cut--off) is slightly smaller
than that of the 3CR sources ($D_{\rm mean} =222 \pm 49$\,kpc; $D_{\rm
med} = 155$\,kpc): the difference falls part--way between the result of
Oort \etal\ \shortcite{oor87}, that the median linear size of radio
sources increases with radio power according to $D_{\rm med} \propto P^m$
where $m = 0.3 \pm 0.05$, and that of Neeser \etal\ \shortcite{nee95}
which showed no significant radio power dependence for radio sizes. The
low number statistics in our small subsamples prohibits any significant
conclusions from being drawn. The good overlap in linear sizes between the
two samples does, however, mean that comparisons of the optical properties
of these sources (Paper II) can safely be made.

\begin{figure}
\centerline{
\psfig{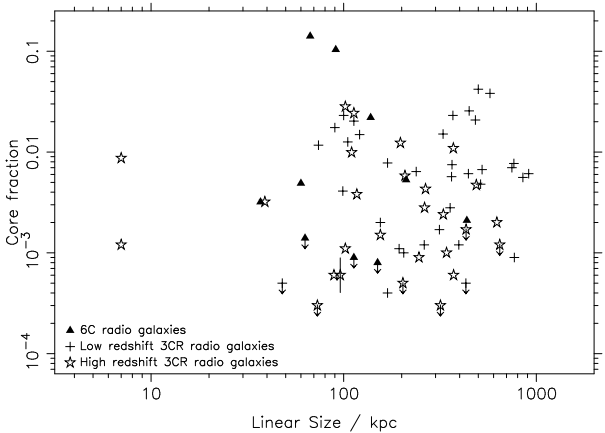}
}
\caption{\label{lincores} A plot of core fraction at a rest--frame of
approximately 16\,GHz versus linear radio size for the 6C radio galaxies
presented here, the $z \sim 1$ 3CR radio galaxies from the sample of Best
\etal\ (1997), and a low redshift ($z < 0.3$) sample of 3CR galaxies from
the sample of Hardcastle \etal\ (1998). See text for more details.}
\end{figure}
\nocite{har98b}

\subsection{Radio core powers}

In Table~\ref{radprops} we tabulate the core fraction, $R$, of the 6C
subsample, defined as the ratio of the flux of the compact central
component to that of the extended radio emission as measured at an
observed frequency of 8\,GHz; ie. $R = f_{\rm core} / (f_{\rm total} -
f_{\rm core})$. With the exception of 1217+36 all of the cores appear
unresolved. Due to the uncertainty in determining the core flux, $1217+36$
is excluded from further analysis.

Core fractions were also determined for the sample of 3CR radio galaxies
with redshifts $z \sim 1$ of Best \etal\ \shortcite{bes97c}\footnote{For
some sources for which the data of Best \etal\ \shortcite{bes97c} provided
only upper limits to the core fractions, values are available from the
literature: 3C324 and 3C368 \cite{bes98a}; 3C49 and 3C241, by
interpolating the core flux densities at 5 and 15\,GHz
\cite{fan89,aku91,bre92} to 8\,GHz; 3C337 \cite{ped89b} has a 5\,GHz core
flux available, from which an 8\,GHz flux can be derived to within a
factor of two, assuming the core spectral index to lie between $-0.5$ and
1 --- in the figures, this point is plotted at the centre of this range,
together with an error bar.}, and a sample of 3CR FR\,II radio galaxies
with redshifts $z < 0.3$ described by Hardcastle \etal\
\shortcite{har98b}. The eight objects classified as broad--line radio
galaxies have been excluded from the latter sample, since the closer
orientation of the radio axes of these objects to the line of sight than
that of narrow--line radio galaxies leads to Doppler boosting of the
central cores, enhancing their $R$ value (e.g. Morganti \etal\ 1997,
Hardcastle \etal\ 1998)\nocite{har98b,mor97}.  For the same reason, 3C22
was excluded from the high redshift 3CR sample
\cite{raw95,eco95}.  Further, to allow a comparison between the low and
high redshift samples, account must be taken of the difference in
rest--frame frequency due to the different spectral indices of the core
and extended emission regions. For this reason, the core fractions of the
low redshift sources have been adjusted to the values that would be
measured at rest--frame 16\,GHz (observed frame 8\,GHz for a source at
redshift $z=1$) under the assumption that the cores have a flat radio
spectrum and the lobes have $\overline{\alpha} \sim 0.8$.

In general, the core fraction of a radio source depends most strongly upon
its orientation, through Doppler boosting effects. Here, however, since we
have begun by selecting samples from low frequency radio surveys, and then
excluded the quasar and broad--line radio galaxy populations, it
has been possible to restrict our analysis to sources oriented close to
the plane of the sky where beaming effects are of little importance.  The
measured core fractions therefore reflect an intrinsic property of the
radio sources.

In Figure~\ref{lincores}, the core fraction is plotted against the linear
size of the radio source for each of the three samples; no significant
correlation is seen. This lack of correlation naively implies a strong
self--similarity in the growth of FR\,II radio sources. Models of the
evolution of FR\,II radio sources through the P--D diagram (e.g. Kaiser
\etal\ 1997 and references therein) \nocite{kai97b} indicate, however,
that as the size of a radio source increases from about 10\,kpc to a few
hundred kpc, the radio luminosity of its lobes will fall by about a factor
of 3. The core flux is not expected to change during this period, thus
resulting in a theoretical increase in the core fraction with linear
size. Figure~\ref{lincores} shows only little evidence for such an
increase; the median core fraction for the sources smaller than 100\,kpc
is 0.0037, increasing to 0.0056 for the sources larger than 300\,kpc. At
any given linear size there is a large scatter in the core fractions, and
so to properly investigate this result it is important to sample a wider
range of linear sizes than is possible here, by adding to the current
diagram a well-defined sample of compact sources and, more importantly, a
sample of giant radio galaxies ($D \gta 1$\,Mpc) where the fall-off of
lobe flux with radio size is expected to be strong.

\begin{figure}
\centerline{
\psfig{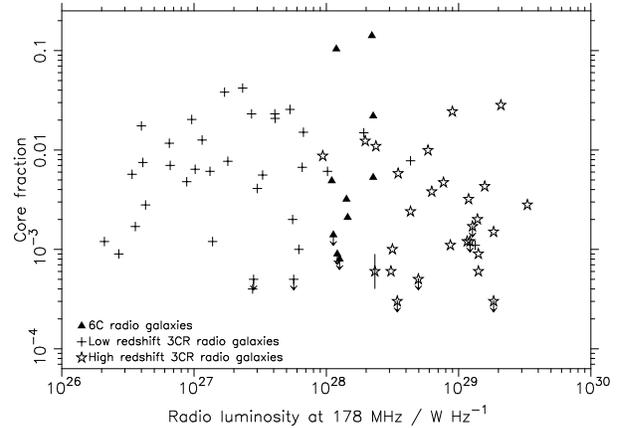}
}
\caption{\label{cores} A plot of 8\,GHz core fraction versus radio
luminosity for the same samples as Figure~\ref{lincores}.}
\end{figure}

Figure~\ref{cores} shows the same core fraction parameter plotted against
the total radio luminosity of the source. There is only a very weak (92\%
significant in a Spearmann Rank test) inverse correlation between these
parameters, over three orders of magnitude in radio luminosity.
Equivalently, for the combined samples we derive a tight correlation
between the core flux and the extended flux, with a best--fitting
power--law function of $P_{\rm core} \propto P_{\rm extended}^{0.74 \pm
0.06}$ (see also e.g. Fabbiano \etal\ 1984, who found $P_{\rm core}
\propto P_{\rm total}^{0.75 \pm 0.05}$ within a sample of low redshift 3CR
FR\,I and FR\,II sources).\nocite{fab84} 

This latter result is particularly interesting when viewed with reference
to the question of why powerful radio sources are so powerful. The
nearby radio galaxy Cygnus A, for example, is approximately 1.5
orders of magnitude more radio luminous than any other nearby FR\,II radio
galaxy, but Barthel and Arnaud \shortcite{bar96} have pointed out that its
far--infrared luminosity, AGN X--ray luminosity and integrated emission
line luminosity are not extreme. They therefore suggest that the
`anomalous' radio loudness of Cygnus A, and other powerful radio sources,
are entirely attributable to their location in a cluster: the denser
surrounding medium reduces the expansion losses of the synchrotron
electron population, resulting in a more efficient transfer of AGN power
into radio emission. 

Whilst this effect is undoubtedly important, if this were the only
responsible process then the radio core emission of Cygnus A and other
powerful radio galaxies should not be enhanced, and their core fractions
should consequently be low. In fact, Cygnus A (one of our low redshift 3CR
sample) can be seen on Figure~\ref{cores} at $P= 1.3 \times 10^{29}$, $R =
0.0011$), in a similar location to the powerful high redshift 3CR
sources. Its core fraction is within the range of values of the other
sources in the low redshift sample, although at the lower end of that; its
core flux density is the highest within that sample. This result means
that the high radio power of Cygnus A must arise primarily due to a high
radio power of its AGN. The dense surrounding environment may further
boost its radio power, resulting in its lower than average core fraction,
but this can not be the dominant effect.

Generalising this result to the whole sample, the implication of the
strong radio core power versus extended power correlation is that the high
radio powers of the most powerful sources must originate in the AGN. The
fact that the coefficient of the correlation power-law is less than unity
($P_{\rm core} \propto P_{\rm extended}^{0.74 \pm 0.06}$), however, leaves
scope for, and indeed requires, an additional effect which is likely to be
environmental.

\subsection{Core spectral indices}

A second interesting feature of the radio cores is their spectral index
properties. At low redshifts, radio source cores generally have flat or
inverted spectra $-1 < \alpha < 0.5$, whilst recent results have shown
that the radio cores in samples of radio galaxies with redshifts $z \gta
2$ often have steep spectra, $\alpha > 0.5$, between observed frequencies
of 5 and 8\,GHz \cite{car97,ath97}. Lonsdale \etal\ \shortcite{lon93} and
Athreya \etal\ \shortcite{ath97} have interpreted this as being due not to
a cosmic epoch effect, but simply to the different rest--frame frequencies
at which the high and low redshift samples are observed. In their models,
the cores of all radio sources are flat due to synchrotron self absorption
at low frequencies, but steepen rapidly above about 20\,GHz, the precise
break frequency varying slightly from source to source.

The 6C sources show a wide range in core spectral indices: in five cases
the spectral index is flat or inverted; the `core' of 1217+36 has an
intermediate spectral index, but contamination from lobe emission suggests
that the true core will be flatter; 1257+36 may have a flat or steep
spectrum core, depending upon which candidate is the true core; 1204+35 is
a clear case of a steep spectrum core. Interestingly, this last source is
also one of the highest redshift objects in the sample ($z=1.37$), and
would fit the picture described above if it was an example of a source
with a below average break frequency.

There is, unfortunately, no large sample of 3CR radio galaxies at this
redshift for which sufficiently deep two--frequency radio observations
have been made to obtain core spectral indices. The few cases for which
these measurements are available have also provided varied results, for
example from the inverted core ($\alpha \approx -0.17$) of 3C49
\cite{bre92,fan89} to a relatively steep spectrum core ($\alpha = 0.54$)
in 3C368 \cite{bes98a}. 

\subsection{Rotation measure properties}

The mean rotation measures determined for the 6C radio galaxies are in no
cases extreme, generally being below 40\,rad\,m$^{-2}$. In many sources,
however, there is a significant difference between the mean values
determined for the two lobes, and in some instances strong gradients are
seen within individual lobes in the depolarisation and rotation measure
maps. Such large variations are unlikely to have their origin in our
galaxy \cite{lea87}, instead being caused by gas in the neighbourhood of
the radio sources, meaning that the rest--frame rotation measures are a
factor of $(1+z)^2$ larger than those quoted in Table~\ref{regprop}.

The large rotation measure differences between the radio lobes of the 6C
sample, 102\,rad\,m$^{-2}$ in the rest--frame on average, are in stark
contrast to the study by Simonetti and Cordes \shortcite{sim86} who found
typically less than 10\,rad\,m$^{-2}$ difference between the two lobes of
low redshift 3C and 4C radio sources. Steep gradients in the
depolarisation and rotation measure are indicative of a dense and clumpy
environment surrounding the radio source, suggesting that the 6C sources
with redshifts $z \sim 1$ lie in a denser environment than those
nearby. An interesting question is whether this increase of environmental
density correlates with radio power or with redshift; Cygnus A, a low
redshift source of comparable radio power to the distant 3CR sources,
shows an extreme range of rotation measures ($-$4000 to
+3000\,rad\,m$^{-2}$; Dreher \etal\ 1987)\nocite{dre87} supporting the
former.

Pedelty \etal\ \shortcite{ped89a} determined the rotation measures of a
sample of 12 high redshift 3CR radio galaxies at somewhat lower angular
resolution (typical beam-size $\sim 1.5''$), and found a mean rest--frame
difference of about 175\,rad\,m$^{-2}$ between the two radio lobes, a
value even more extreme than that of the current 6C sample. They also
observed structure within the lobes, particularly in the case of 3C337
\cite{ped89b}. Johnson \etal\ \shortcite{joh95} selected three high
redshift 3CR radio sources with angular scales in excess of 40 arcsecs,
and also found considerable rotation measure structure within individual
lobes. Best \etal\ \shortcite{bes98a} studied two radio galaxies from the
Pedelty \etal\ sample, 3C324 and 3C368, with the same high angular
resolution obtained for the observations presented in this paper, and
found considerable structure on scales smaller than those previously
probed. Gradients of up to 1000\,rad\,m$^{-2}$ over distances of about
10\,kpc were measured. If the results from these small samples are
representative of the $z \sim 1$ 3CR sources, then it would suggest that
the distant 3CR sources live in still denser environments than the 6C
sources at the same redshift.

\subsection{Separation Quotients}
\label{sepquo}

The separation quotient, $Q$, is defined as the ratio $Q =
\theta_1/\theta_2$, where $\theta_1$ and $\theta_2$ are the angular
distances from the nucleus of the more distant and closer hot-spots
respectively \cite{ryl67,lon79a}. These arm--lengths are tabulated in
Table~\ref{regprop}, and the separation quotients in
Table~\ref{radprops}. The 6C galaxies show considerable asymmetries, with
a mean value of $\overline{Q} = 1.80 \pm 0.28$, although this is somewhat
dominated by the very high asymmetries of 0943+39 and 1017+37; the median
value of $Q$ for the sample is 1.47. This can be compared to 3CR radio
galaxies in the same redshift range which have $\overline{Q} = 1.39 \pm
0.07$, with a median value of 1.37 (data taken from Best et~al. 1995,
1997). The 6C galaxies appear to show higher asymmetry quotients, although
to obtain statistically significant confirmation study of a larger sample of
lower power sources at these redshifts, such as those in the Molonglo
Strip \cite{mcc97} will be required.

The separation quotient has two origins. Firstly, the light--travel time
differences from the two hotspots for any source which is not orientated
precisely in the plane of the sky will give rise to an apparent asymmetry:
this has been used by a number of authors to investigate hotspot advance
speeds \cite{lon79a,ban80,bes95a,sch95}. Secondly, environmental effects
may produce intrinsic asymmetries (e.g. McCarthy \etal
1991)\nocite{mcc91}. Best et~al \shortcite{bes95a} compared radio galaxies
and quasars using this method to test orientation--based unification
schemes \cite{bar89}, and suggested that for radio galaxies the two effects
are roughly comparable. In this respect, a higher separation quotient for
the 6C sources at these redshifts would be an interesting result, possibly
reflecting a greater environmental influence on the lower power jets of
the 6C sources.

\section{Conclusions}
\label{concs}

We have presented total intensity, spectral index, polarisation and
rotation measure radio maps of a complete sample of eleven galaxies from
the 6C catalogue, together with infrared images of the fields surrounding
them. Basic source parameters were also tabulated. The data were compared
with a sample of more powerful 3CR radio galaxies at the same redshift,
and with a low redshift radio galaxy sample. The main results can be
summarized as follows.

\begin{itemize}
\item All of the sources display an FR\,II type morphology, with the
possible exception of 1217+36 in which the compact central double
component is surrounded by an extended halo of radio emission. Most of the
sources, however, show some deviation from the `standard double'
morphology, either as double hotspots or hotspots lying withdrawn from the
leading edge of the lobe.

\item We have detected radio cores in eight of the eleven sources, four
for the first time, co-incident with the optical identifications. In at
least two, and possibly as many as five cases, the core is inverted. In
one case the core has a steep spectrum ($\alpha \sim 0.9$), consistent
with the trend for radio cores at high redshifts to high steeper spectral
indices \cite{car97,ath97}.

\item The ratio of radio luminosity of the core to that of the extended
emission, $R$, appears to increase less rapidly with the linear size of
the radio source than predicted by radio source evolution models. It is
only weakly anti--correlated with the total radio power, implying that the
total radio power of a radio source is determined primarily by the
AGN. Environmental effects play a secondary role.

\item The rotation measures detected are in no cases extreme (generally
averaging below 40\,rad\,m$^{-2}$), but strong gradients are seen in both
the depolarisation and rotation measures between the two lobes and often
within individual lobes. These gradients are significantly larger than
those of low redshift radio sources, suggesting than the 6C sources live
in a dense, clumpy environment. On the other hand, they appear lower than
those of the 3CR sources at the same redshift, possibly indicating a
weak dependence of radio power on local environmental density.
\end{itemize}

\section*{Acknowledgements} 

The National Radio Astronomy Observatory is operated by Associated
Universities Inc., under co-operative agreement with the National Science
Foundation.  This work was supported in part by the Formation and
Evolution of Galaxies network set up by the European Commission under
contract ERB FMRX-- CT96--086 of its TMR programme. We thank the referee,
Dr T.W.B. Muxlow, for useful comments.

\label{lastpage}
\bibliography{pnb} 
\bibliographystyle{mn} 

\end{document}